\documentclass[twocolumn,showpacs,preprintnumbers,amsmath,amssymb,a4paper,superscriptaddress, nofootinbib]{revtex4-1}

\usepackage{amsmath}
\usepackage{graphicx}
\usepackage{dcolumn}
\usepackage{bm}

\usepackage{latexsym}
\usepackage{amssymb}
\usepackage{epstopdf}
\usepackage{eqnarray}

\begin{document}
\title{Anisotropic quantum scattering in two dimensions}
%31.15.ac High-precision calculations for few-electron (or few-body) atomic systems
%31.15.xf   Finite-difference schemes
%34.50.Cx        Elastic; ultracold collisions
%34.50.-s   Scattering of atoms and molecules
\author{Eugene A. Koval}
\email[]{e-cov@yandex.ru}
\affiliation{Bogoliubov Laboratory of Theoretical Physics, Joint Institute for Nuclear Research, Dubna, Moscow Region 141980, Russian Federation}
\affiliation{Department of Theoretical Physics, Dubna International University for Nature, Society and Man, Dubna, Moscow Region 141980, Russian Federation}

\author{Oksana A. Koval}
\email[]{kov.oksana20@gmail.com}
\affiliation{Bogoliubov Laboratory of Theoretical Physics, Joint Institute for Nuclear Research, Dubna, Moscow Region 141980, Russian Federation}

\author{Vladimir S. Melezhik}
\email[]{melezhik@theor.jinr.ru}
\affiliation{Bogoliubov Laboratory of Theoretical Physics, Joint Institute for Nuclear Research, Dubna, Moscow Region 141980, Russian Federation}%
\affiliation{Department of Theoretical Physics, Dubna International University for Nature, Society and Man, Dubna, Moscow Region 141980, Russian Federation}

\date{\today}

\begin{abstract}\label{txt:abstract}
We study the quantum scattering in two spatial dimensions (2D)
without the usual partial-wave formalism. The analysis beyond the
partial-wave approximation allows a quantitative treatment of the
anisotropic scattering with a strong coupling of different angular
momenta nonvanishing even at the zero-energy limit. High
efficiency of our method is demonstrated for the 2D scattering on
the cylindrical potential with the elliptical base and
dipole-dipole collisions in the plane. We reproduce the result for
the 2D scattering of polarized dipoles in binary collisions
obtained recently by Ticknor [Phys. Rev. A {\bf 84}, 032702
(2011)] and explore the 2D collisions of unpolarized dipoles.
\end{abstract}

\maketitle

\section{Introduction}
In recent years, the problem of anisotropic quantum scattering in
two spatial dimensions (2D) attracts increasing interest. It is
stimulated by the spectacular proposals for prospects to create
exotic and highly correlated quantum systems with dipolar gases
\cite{ref1,ref2}. Particularly, there were considered anisotropic
superfluidity \cite{ref3}, 2D dipolar fermions \cite{ref4}, and
few-body dipolar complexes \cite{ref5}. The recent experimental
production of ultracold polar molecules in the confined geometry
of optical traps \cite{ref6,ref7,ref8} has opened up ways to
realize these phenomena. Noteworthy also is a rather long history
of research of 2D quantum effects in condensed matter physics. One
can note superfluid films \cite{ref9}, high-temperature
superconductivity \cite{ref10}, 2D materials, such as graphene
\cite{ref11}, and even possibilities for topological quantum
computation \cite{ref12}. Unique opportunities for modeling these
2D effects in a highly controlled environment have recently
appeared with the development of experimental techniques for
creating quasi-2D Bose and Fermi ultracold gases \cite{ref13}.

Interest in the processes and effects in 2D-geometry has
stimulated the theory of elementary quantum two-body systems and
processes in the plane. Special consideration should be given to
the anisotropy and long-range character of the dipole-dipole
interaction. Actually, usual partial-wave analysis becomes
inefficient for describing the dipole-dipole scattering due to the
strong anisotropic coupling of different partial-waves in the
asymptotic region \cite{ref14,ref15}. Recently, considerable
progress in the analysis of the 2D and quasi-2D (q2D) scattering
of dipoles has been achieved~\cite{ref16,ref17,ref18,ref19,ref20}.
Thus, the 2D dipolar scattering in the threshold and semiclassical
regimes was studied in the case of the dipole polarization
directed orthogonally to the scattering plane~\cite{ref16}. An
arbitrary angle of polarization was considered in \cite{ref17}.

In this work, we develop a method for quantitative analysis of the 2D quantum
scattering on a long-range strongly anisotropic scatterer. Particularly, it
permits the description of the 2D collisions of unpolarized dipoles. Our
approach is based on the method suggested in \cite{ref21} for the few-dimensional
scattering which was successfully applied to the dipole-dipole scattering
induced by an elliptically polarized laser field in the 3D free-space \cite{ref15}.

The key elements of the method are described in Section II. In
Section III, we apply the method to the 2D scattering on the
cylindrical potential with the elliptical base and the 2D
dipole-dipole scattering of unpolarized dipoles. We reproduce the
threshold formula \cite{ref22b,ref24} for the scattering amplitude
on the cylinder potential with the circular base and the results
of \cite{ref16,ref17} for the 2D scattering of polarized dipoles.
High efficiency of the method has been found in all problems being
considered. The last Section contains the concluding remarks. Some
important details of the computational scheme and illustration of
the convergence are given in Appendices.

 \section{2D Scattering problem in angular-grid representation}
 The quantum scattering on the anisotropic potential $U(\rho ,\phi )$ in
 the plane is described by the 2D Schr\"{o}dinger equation in polar
 coordinates $(\rho ,\phi )$
 \begin{equation}
 \label{eq1}
 H(\rho ,\phi )\Psi \left( {\rho ,\phi } \right)=E\Psi \left( {\rho
 ,\phi } \right)
 \end{equation}
 with the scattering boundary conditions
 \begin{equation}
 \label{eq2}
 \Psi \left( {\rho ,\phi } \right)\to e^{i {\bm q}\bm{\rho}
  }+f\left( {q,\phi ,\phi _q }
 \right)\frac{e^{iq\rho }}{\sqrt {-i\rho } }
 \end{equation}
 in the asymptotic region $\rho \to \infty $ and the Hamiltonian of the
 system
 \[
 H(\rho ,\phi )=-\frac{\hbar ^2}{2\mu }\left( {\frac{1}{\rho
 }\frac{\partial }{\partial \rho }\left( {\rho \frac{\partial }{\partial \rho
 }} \right)+\frac{1}{\rho ^2}h^{(0)}(\phi )} \right)+U\left( {\rho
 ,\phi } \right).
 \]
 The unknown wave function $\Psi \left( {\rho ,\phi } \right)$ and the
 scattering amplitude $f\left( {q,\phi ,\phi _q } \right)$ are searched
 for the fixed momentum ${\bm q}$ defined by the colliding energy $E$
 ($q= \sqrt {2\mu E}/\hbar )$ and the direction ${\bm q}/q$ of the incident wave (defined by the angle
 $\phi _q )$ and for the scattering angle $\phi $ \footnote{ Hereafter
 we use the definition of the scattering amplitude introduced in \cite{ref22a}.}. Here
 $\mu $ is the reduced mass of the system. In the polar coordinates, the
 angular part of the kinetic energy operator in $H(\rho ,\phi )$ has a
 simple form $h^{(0)}(\phi )=\frac{\partial ^2}{\partial \phi ^2}$. The
 interaction potential $U\left( {\rho ,\phi } \right)$ can be anisotropic in the general
 case, i.e. to be strongly dependent on $\phi $. It is clear
 that varying the direction of the incident wave ${\bm q}/q$ can be replaced by the rotation $U\left( {\rho
 ,\phi } \right)\to U\left( {\rho ,\phi +\phi _q } \right)$ of the
 interaction potential by the angle $\phi _q $ for the fixed direction of
 the incident wave, which we choose to be coincident with the x-axis. Thus, in the case
 of anisotropic potential $U\left( {\rho ,\phi } \right)$ the task is to
 solve the problem (\ref{eq1}) with the interaction potential $U\left( {\rho ,\phi
 +\phi _q } \right)$ for all possible $\phi _q $ and fixed $E$ with the
 scattering boundary conditions
 \begin{equation}
 \label{eq3}
 \Psi \left( {\rho ,\phi } \right)\to \exp \{iq\rho \cos (\phi
 )\}+f\left( {q,\phi ,\phi _q } \right)\frac{e^{iq\rho }}{\sqrt {-i\rho
 } }\,.
 \end{equation}
 If the scattering amplitude $f\left( {q,\phi ,\phi _q } \right)$ is
 found, one can calculate the differential scattering cross section
 \begin{equation}
 \label{eq4}
 {d\sigma (q,\phi,\phi_q )} \mathord{\left/ {\vphantom {{d\sigma (q,\Omega )}
 {d\Omega }}} \right. \kern-\nulldelimiterspace} {d\Omega
 }\mbox{=}{\left| {f(q,\phi ,\phi
 _q )} \right|^2}\,,
 \end{equation}
 where $d\Omega = d\phi d\phi_q$, as well as the total cross section
 \begin{equation}
 \label{eq5}
 \sigma (q)\mbox{=}\frac{1}{2\pi }\int\limits_0^{2\pi }
 \int\limits_0^{2\pi } {\frac{d\sigma }{d\Omega }}
d\phi _q d\phi
 \end{equation}
 by averaging over all possible orientations $\phi _q $ of the scatterer
 and integration over the scattering angle $\phi $.

 To integrate the problem
 %(\ref{eq1}),(\ref{eq2})
(\ref{eq1}),(\ref{eq2}),
 we use the method suggested in \cite{ref21} to
 solving a few-dimensional scattering problem and applied in \cite{ref15} for the
 dipole-dipole scattering in the 3D free-space. Following the ideas of these
 works we choose the eigenfunctions
 \begin{equation}
 \label{eq6}
 \xi _m (\phi )=\frac{1}{\sqrt {2\pi } }e^{im\left( {\phi -\pi }
 \right)}=\frac{(-1)^m}{\sqrt {2\pi } }e^{im\phi }
 \end{equation}
 of the operator $h^{(0)}(\phi )$ as a Fourier basis for the angular-grid
 representation of the searched wave-function $\Psi \left( {\rho ,\phi }
 \right)$. We introduce the uniform grid $\phi _j =\frac{2\pi j}{2M
 +1}\mbox{ }(\mbox{where }j=0,1,...,2M$) over the $\phi $ and $\phi _q $-variables and search the wave
 function as expansion
 \begin{equation}
 \label{eq7}
 \begin{array}{l}
  \Psi (\rho ,\phi )=\frac{1}{\sqrt \rho }\sum\limits_{j=0}^{2M }
 {\sum\limits_{m=-M}^{M} {\xi _m
 (\phi )\xi _{mj}^{-1} \psi _j (\rho )} } = \\
  =\frac{2\pi }{\sqrt \rho }\left( {\frac{1}{2M +1}}
 \right)\sum\limits_{j=0}^{2M } {\sum\limits_{m=-M}^{M} {e^{im(\phi -\phi _j )}\psi _j (\rho )}
 } \,, \\
  \end{array}
 \end{equation}
 where $\xi _{mj}^{-1} =\frac{2\pi }{2M +1}\xi _{jm}^\ast =\frac{2\pi
 }{2M +1}e^{-im\phi _j }$ is the inverse matrix to the $\left(
 {2M +1} \right)\times \left( {2M +1} \right)$ square matrix
 $\xi _{jm} =\xi _m (\phi _j )$ defined on the angular grid\footnote{ To
 calculate the inverse matrix $\xi _{mj}^{-1} $, we use the completeness
 relation for the Fourier basis $\sum\limits_{m=-\infty }^\infty {\xi _m
 (\phi _k )\xi _m^\ast (\phi _j )=\delta (\phi _k -\phi _j )} $,
 which in our grid representation reads $\sum\limits_{m=-M}^{M} {\xi _{km} \xi _{jm}^\ast =\frac{2M
 +1}{2\pi }\delta _{kj} } $.}.

 In the representation (\ref{eq7}) the unknown coefficients $\psi _j (\rho )$ are defined
 by the values of the searched wave function on the angular grid $\psi _j
 (\rho )=\sqrt \rho \Psi (\rho ,\phi _j )$, any local interaction is
 diagonal
%\begin{align}
% U(\rho ,\phi )\Psi (\rho ,\phi )\left| {_{\phi =\phi _j } } \right. =\nonumber \\
%=U(\rho ,\phi _j )\frac{1}{\sqrt \rho }\frac{2\pi }{(2M +1)} \sum\limits_{k=0}^{2M } {\sum\limits_{m=-M}^{M} {e^{im(\phi _j -\phi _k )}\psi _k (\rho)} } & \nonumber \\
%=\frac{U(\rho ,\phi _j )}{\sqrt \rho }\psi _j (\rho ),
%\label{eq8}
%\end{align}

%\begin{align}
% U(\rho ,\phi )\Psi (\rho ,\phi )\left| {_{\phi =\phi _j } } \right. &=\nonumber \\
%&=U(\rho ,\phi _j )\frac{1}{\sqrt \rho }\frac{2\pi }{(2M
% +1)} \nonumber \\
% &\qquad {} \times \sum\limits_{k=0}^{2M } {\sum\limits_{m=-M
% }^{M} {e^{im(\phi _j -\phi _k )}\psi _k (\rho
% )} } \nonumber \\
% &=\frac{U(\rho ,\phi _j )}{\sqrt \rho }\psi _j (\rho ),
%\label{eq8}
%\end{align}

\begin{equation}
\begin{array}{lr}
 U(\rho ,\phi )\Psi (\rho ,\phi )\left| {_{\phi =\phi _j } } \right. =\\
= \frac{2\pi }{(2M+1)\sqrt{\rho}}U(\rho ,\phi _j
)\sum\limits_{j'=0}^{2M } {\sum\limits_{m=-M}^{M} {e^{im(\phi _j
-\phi _{j'} )}\psi _{j'} (\rho
 )} } \\
=\frac{1}{\sqrt \rho }U(\rho ,\phi _j )\psi _j (\rho )\,,
\label{eq8}
\end{array}
\end{equation}

% \begin{equation}
% \label{eq8}
% U(\rho ,\phi )\Psi (\rho ,\phi )\left| {_{\phi =\phi _j } }
% \right.=U(\rho ,\phi _j )\frac{1}{\sqrt \rho }\frac{2\pi }{(2M
% +1)}\sum\limits_{k=0}^{2M } {\sum\limits_{m=-M
% }^{M} {e^{im(\phi _j -\phi _k )}\psi _k (\rho
% )} } =\frac{U(\rho ,\phi _j )}{\sqrt \rho }\psi _j (\rho ),
% \end{equation}
\noindent and the angular part $h_{jj'}^{(0)} $ of the kinetic energy operator has a
 simple form
% \begin{align}
%  \label{eq9}
% h^{(0)}(\phi )\Psi (\rho ,\phi )\left| {_{\phi =\phi _j } }
% \right.=\frac{1}{\sqrt \rho }\sum\limits_{k=0}^{2M } {h_{jk}^{(0)}
% \Psi _k (\rho )} \nonumber \\
% =-\frac{2\pi }{(2M +1)\sqrt \rho
% }\sum\limits_{m=-M}^{M}
% {\sum\limits_{k=0}^{2M } {m^2e^{im(\phi _j -\phi _k )}\psi _k
% (\rho )} } .
% \end{align}

 \begin{equation}
  \label{eq9}
 \begin{array}{l}

 h^{(0)}(\phi )\Psi (\rho ,\phi )\left| {_{\phi =\phi _j } }
 \right.=\frac{1}{\sqrt \rho }\sum\limits_{j'=0}^{2M } {h_{jj'}^{(0)}
 \psi _{j'}(\rho )} \\
 =-\frac{2\pi }{(2M +1)\sqrt \rho
 }\sum\limits_{j'=0}^{2M } (
 {\sum\limits_{m=-M}^{M} {m^2e^{im(\phi _j -\phi _{j'} )})\psi _{j'}
 (\rho )} }\,.
 \end{array}
 \end{equation}
 Note that the presence in the interaction potential of the
 ``nonlocal'' angular part (i.e. the integration or differentiation over
 angular variable) leads to destroying the diagonal structure in (\ref{eq8}).

 Thus, the 2D Schr\"{o}dinger equation (\ref{eq1}) is reduced in the angular-grid
 representation (\ref{eq7}) to the system of \mbox{$2M +1$} coupled ordinary
 differential equations of the second order:
\begin{align}
 \label{eq10}
 \frac{d^2\psi _j (\rho )}{d\rho ^2}+\frac{2\mu }{\hbar ^2}\left( {E-U(\rho
 ,\phi _j )
 +\frac{\hbar ^2}{8\mu \rho ^2}} \right)\psi _j (\rho
 )+\nonumber \\
+\frac{1}{\rho ^2}\sum\limits_{j'} {h_{jj'}^{(0)} \psi _{j'} (\rho
)} =0\,.
\end{align}
% \begin{equation}
%
% \label{eq10}
% \begin{array}{l}
% \frac{d^2\psi _j (\rho )}{d\rho ^2}+\frac{2\mu }{\hbar ^2}\left( {E-U(\rho
% ,\phi _j )
% +\frac{\hbar ^2}{8\mu \rho ^2}} \right)\psi _j (\rho
% )+\\
%+\frac{1}{\rho ^2}\sum\limits_{j'} {h_{jj'}^{(0)} \psi _{j'} (\rho )} =0.
%  \end{array}
% \end{equation}
 Since the wave function $\Psi (\rho ,\phi _j )=\frac{\psi _j (\rho
 )}{\sqrt \rho }$ must be finite at the origin $\left( {\frac{\psi _j (\rho
 )}{\sqrt \rho }\to const} \right)$, the ``left-side'' boundary condition for
 the functions $\psi _j (\rho )$ reads as
 \begin{equation}
 \label{eq11}
% \psi _j (0)=0\mbox{ }(j=0,1,\ldots ,2M ).
 \psi _j (\rho \to 0)\to const{\sqrt \rho }\quad (j=0,1,\ldots ,2M )\,.
 \end{equation}
 In the asymptotic region $\rho \to \infty $ the scattering boundary
 condition (\ref{eq3}) accepts the form
 \begin{align}
 \label{eq12}
 \frac{2\pi }{\sqrt \rho (2M +1)}\sum\limits_{j=0}^{2M }
 {\sum\limits_{m=-M}^{M}
 {e^{im(\phi -\phi _j )}\psi _j (\rho )} } =\nonumber \\=\exp\{iq\rho \cos (\phi
 )\}+f(q,\phi ,\phi _q )\frac{e^{iq\rho }}{\sqrt {-i\rho } }\,.
 \end{align}

% \begin{equation}
% \label{eq12}
% \frac{2\pi }{\sqrt \rho (2M +1)}\sum\limits_{j=0}^{2M }
% {\sum\limits_{m=-M}^{M}
% {e^{im(\phi -\phi _j )}\psi _j (\rho )} } =e^{iq\rho \cos (\phi
% )}+f(q,\phi ,\phi _q )\frac{e^{iq\rho }}{\sqrt {-i\rho } }
% \end{equation}
\noindent By using the Fourier expansion
 for the plane wave $\exp \{iq\rho \cos (\phi )\}$ and the scattering
 amplitude $f(q,\phi ,\phi_q)$\footnote{ Here $J_m (q,\rho )$ are the first kind
 Bessel functions of integer order. Their asymptotic behavior \cite{ref23}:
 %\[J_m
 %(z)\xrightarrow[{z\to 0}]{}\left( {\frac{z}{2}} \right)^m\frac{1}{\Gamma
 %(m+1)},\mbox{ m}\ne -1,-2,-3,...\]
\begin{align*}
 J_m (z)\xrightarrow[{\left| z
 \right|\to \infty }]{}\sqrt {\frac{2}{\pi z}} \cos \left( {z-\frac{m\pi
 }{2}-\frac{\pi }{4}} \right)+e^{\left| {Imz} \right|}{\rm O}\left( {\left| z
 \right|^{-1}} \right),\\
 \left( {\left| {\arg (z)} \right|<\pi }
 \right)
\end{align*}
% \[J_m (z)\xrightarrow[{\left| z
% \right|\to \infty }]{}\sqrt {\frac{2}{\pi z}} \cos \left( {z-\frac{m\pi
% }{2}-\frac{\pi }{4}} \right)+e^{\left| {Imz} \right|}{\rm O}\left( {\left| z
% \right|^{-1}} \right),\mbox{ }\left( {\left| {\arg (z)} \right|<\pi }
% \right)\]
 }
 \begin{equation}
 \label{eq13}
 \exp \{iq\rho \cos (\phi )\}=\sum\limits_{m'=-M }^M
 {i^{m'}J_{m'} (q,\rho )e^{i{m}'\phi }}
 \end{equation}
 \begin{equation}
 \label{eq14}
 f(q,\phi ,\phi _q)=\frac{1}{\sqrt {2\pi }
 }\sum\limits_{m'=-M
 }^M {f_{m'}(\phi_q) e^{i{m}'\phi }}
 \end{equation}
 we eliminate the angular dependence from the asymptotic equation (\ref{eq12}) and
 represent the ``right-side'' boundary condition for the functions $\psi _j
 (\rho \to \infty )$ in the form
 \begin{align}
 \label{eq15}
 \frac{2\pi }{(2M +1)\sqrt \rho }\sum\limits_{j=0}^{2M }
 {e^{-im\phi _j }\psi _j (\rho )} =i^mJ_m (q\rho )\sqrt {2\pi }
 + \nonumber \\
 +\frac{f_m(\phi_q)}{\sqrt {-i\rho } }e^{iq\rho }\,.
 \end{align}
 To solve the boundary-value problem
% (\ref{eq10}),(\ref{eq11}),(\ref{eq15})
  (\ref{eq10}),(\ref{eq11}) and (\ref{eq15}),
 we introduce the grid over the $\rho \mbox{-variable} \quad \{\rho _n \}\mbox{
 (}n=0,1,\ldots ,N\mbox{)}$ and reduce the system of differential equations
 (\ref{eq10}) by using the finite-difference approximation of the sixth order to the
 system of $(N+1)\times (2M +1)$ algebraic equations

 \begin{equation}
 \label{my_eq15}
% \hat{O}\vec{\psi} =0
 \hat{A}{\bm \psi} =0
 \end{equation}

\noindent with the band-structure of the matrix $\hat{A}$ with the
width $(2M+1)\times 7$ of the band. By using the asymptotic
equations (\ref{eq15}) in the last
 two points $\rho _{N-1} $ and $\rho _{N}$ one can eliminate the unknown
 vector $f_m(\phi_q)$ from equation (\ref{eq15}) and rewrite the ``right-side''
 boundary condition in the form
\begin{align}
 \label{my_eq16}
 \sum\limits_{j'} \{{A_{j{j}',NN-1} \psi _{{j}'} (\rho_{N-1})} +
 {A_{j{j}',NN} \psi _{{j}'} (\rho _{N} )}\}= \nonumber \\
 = F_{j,N}(q,\rho _{N-1},\rho _{N} )\,.
 \end{align}
 Analogously, one can eliminate unknown constant from expression (\ref{eq11}) by considering asymptotic equations
 (\ref{eq11}) at the first points  $\rho _0 \,, \rho_1$ and $\rho _1$. The acquired ``left-side'' boundary condition reads
 \begin{align}
  \label{my_eq17}
  \psi_j(\rho_0=0)=0\nonumber \,\,, \\
 \sum\limits_{j'}\{{A_{j{j}',11} \psi _{{j}'} (\rho _1)} +
 {A_{j{j}',12} \psi _{{j}'} (\rho _{2} )}\} = F_{j,1}(q,\rho _1 ,\rho _{2} )\,.
 \end{align}

 Thus, the scattering problem is reduced to the boundary value problem
 (\ref{my_eq15}-\ref{my_eq17})
\begin{equation}
 \label{my_eq19}
\hat{A}{\bm \psi} = {\bm F}\,,
\end{equation}
 which can be efficiently solved with standard computational techniques such
 as the sweeping method \cite{ref25} or the LU-decomposition \cite{ref26}. The detailed structure of the matrix of the
 coefficients $A_{jj',nn'} $ is discussed in Appendix A. After the
 solving of Eq.(\ref{my_eq19}) and finding the wave function
 $\psi _j(\rho)$ the scattering amplitude $f(q,\phi,\phi_q)$ is
 constructed according to Eqs.(\ref{eq15}) and (\ref{eq14}).

 \section{Results and Discussion}
 \label{sec:results}
 \subsection{Scattering on anisotropic scatterer}
 First, we have analyzed the 2D scattering on the cylindrical potential barrier
 with the elliptical base
 \begin{equation}
 \label{eq16}
 U(\rho ,\phi )=\left\{ {{\begin{array}{*{20}c}
  {U_0 ,\mbox{ }\rho \leqslant a(\phi )} \hfill \\
  {0,\mbox{ }\rho > a(\phi )}\,\,. \hfill \\
 \end{array} }} \right.
 \end{equation}

\begin{figure}[hbtp]
 \begin{minipage}[hbtp]{0.99\linewidth}
 \center{\includegraphics[width=1\linewidth]{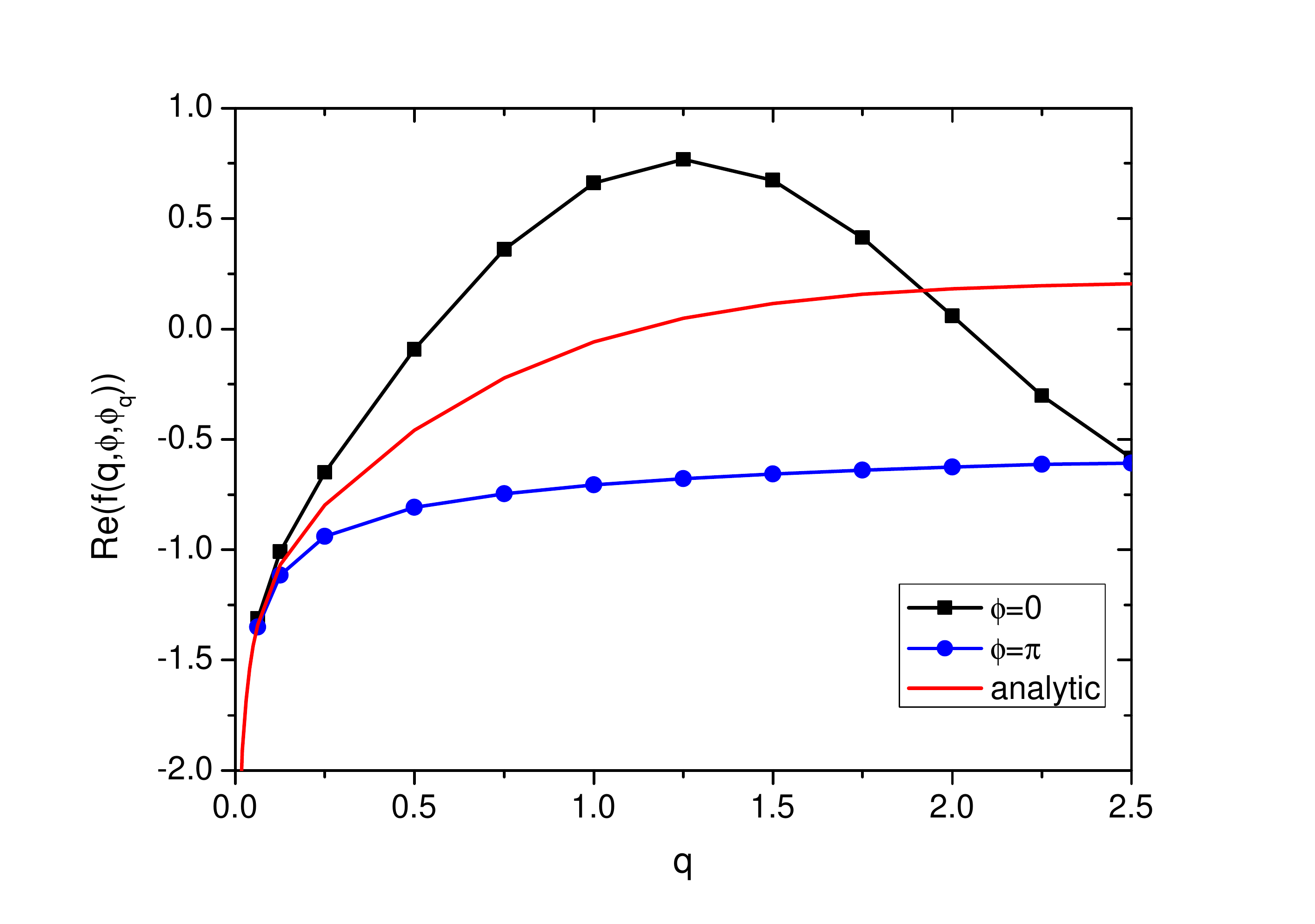} \\ a)}
 \end{minipage}
 \hfill
 \begin{minipage}[hbtp]{0.99\linewidth}
 \center{\includegraphics[width=1\linewidth]{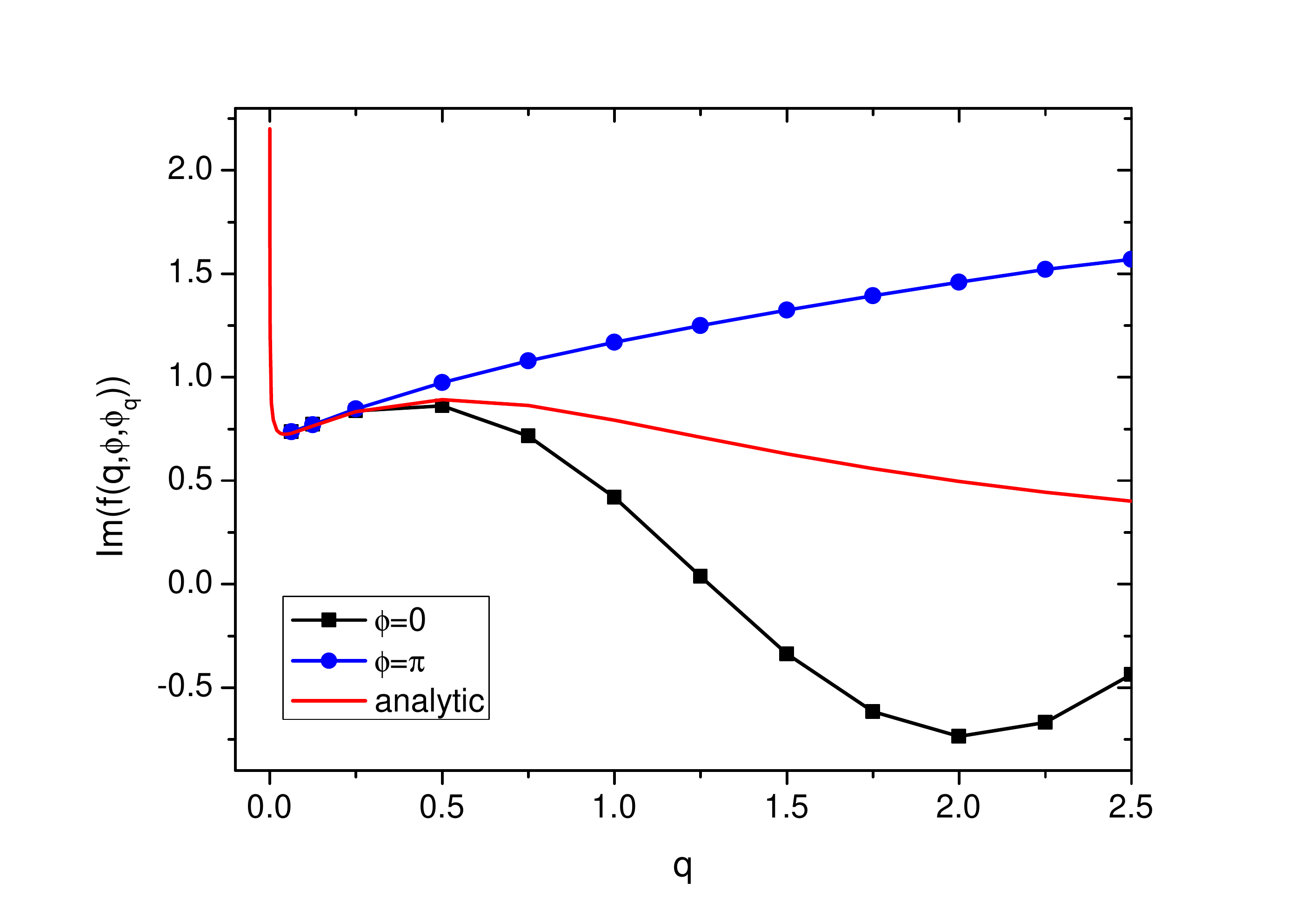} \\ b)}
 \end{minipage}
 \caption{(Color online) The dependence of real a) and imaginary b) part of the scattering amplitude
 on the momentum $q$ for the potential barrier (\ref{eq16}) with circular base $a(\phi)= a_0 =1$ and $U_0=10^4$.
 Calculations were carried out for $\phi_q=0$ with the parameters:
 $M=10^2$, $N=10^5$ and $\rho_N=15$. The quantities $f$ and $q$ are given in the units $\hbar=\mu=1$.}
 \label{fig1}
 \end{figure}

\begin{figure}[hbtp]
 %\centerline{\includegraphics[width=8.80in,height=6.88in]{Pure3.pdf}}
 \centerline{\includegraphics[width=0.99\linewidth]{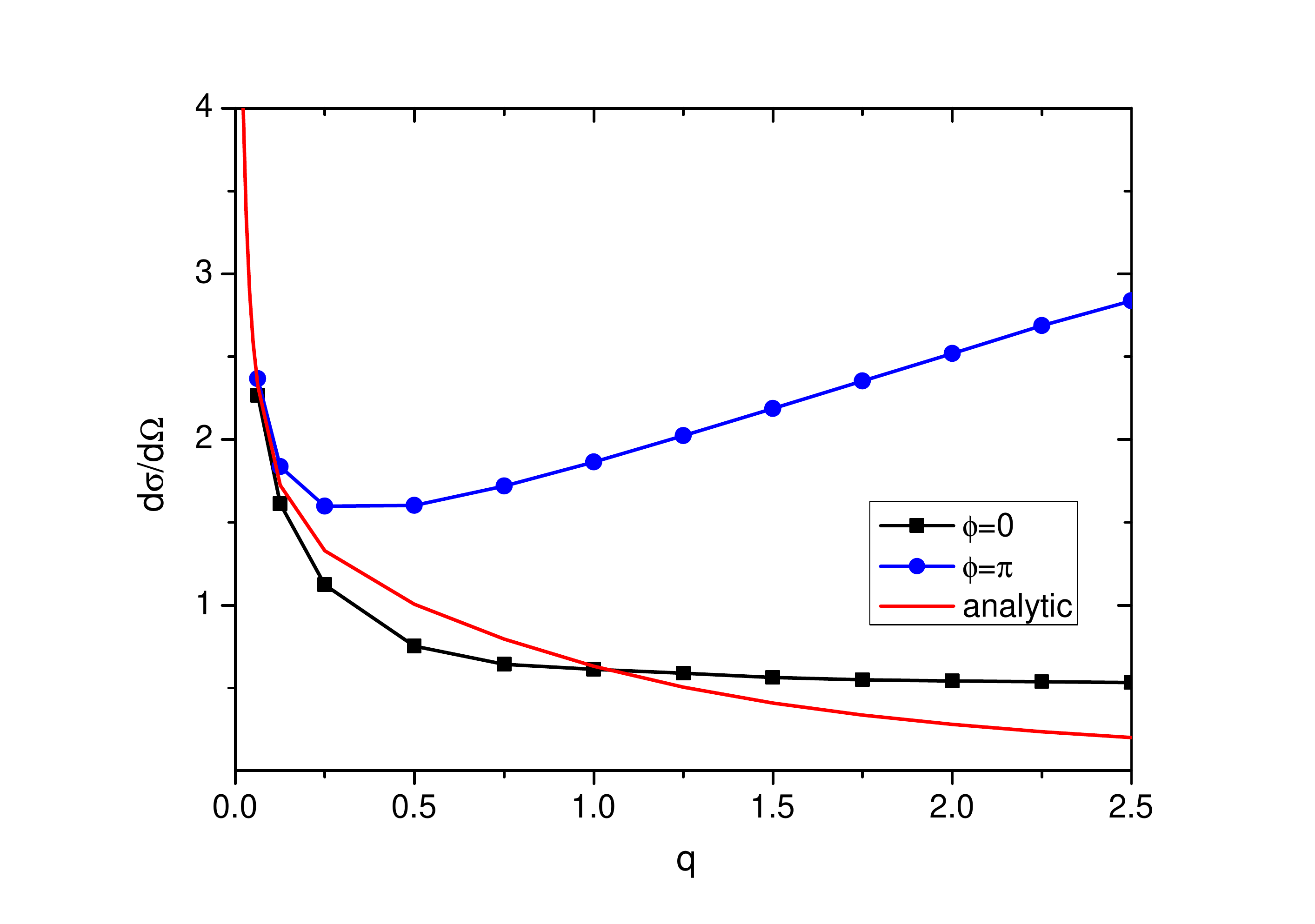}}
 \caption{(Color online) The dependence of the differential cross section on the momentum
 $q$ for the potential barrier (\ref{eq16}) with circular base $a(\phi)= a_0 =1$ and $U_0=10^4$. Calculations were carried out for $\phi_q=0$ with the parameters:
 $M=10^2$, $N=10^5$ and $\rho_N=15$. The quantities $d\sigma/d\Omega$ and $q$ are given in the units $\hbar=\mu=1$.}
 \label{fig3}
 \end{figure}

\noindent The case of the circular base $a(\phi )=a_0 $ was
considered in
 \cite{ref22b,ref24}, where analytic formula for the scattering amplitude
 \begin{equation}
 \label{eq17}
 f(q)\to -\sqrt {\frac{\pi }{2q}} \frac{1}{\ln \left[ {\frac{2}{\gamma q a_{\textrm{2D}}
 }} \right]+i\frac{\pi }{2}}
 \end{equation}
 was obtained at the zero-energy limit $q\to 0$. Here \mbox{$\gamma =\exp
 (C)$} and $C=\mbox{0.577...}$ is the Euler constant. We have analyzed the
scattering on the potential barrier with circular base $a(\phi
)=a_0 $  for arbitrary momentum $q$. The results of calculation
presented in Figs.~\ref{fig1} and \ref{fig3} confirm the
convergence of the scattering amplitude $f(q,\phi,\phi_q)$ to the
analytical value (\ref{eq17}) at $q\rightarrow 0$. In this
Subsection all the calculations were performed in the units
$\hbar=\mu=1$.

 %We have to note that the
 %zero-energy limit (\ref{eq17}) is defined only by the range $\rho _0 $ of the
 %scattering potential. This fact is illustrated in
 %Figs.~\ref{fig1} and \ref{fig2} by
 %comparing the asymptotical values (\ref{eq17}) with the numerical results. In
 %the zero-energy limit $q\to 0$ the scattering amplitude becomes isotropic
 %and coincides with the asymptotic analytic result (\ref{eq17}).

 In the limiting case of the infinitely high potential
 barrier (\ref{eq16}) with the circular base $a(\phi )=a_0$ the
 asymptotic formula (\ref{eq17}) becomes exact for arbitrary $q$. This is confirmed by
 investigation presented in Table~\ref{tab1} which illustrates the convergence of the numerical values
$f(q,\phi,\phi_q )$ with
 increasing ($U_0 \to \infty $) and narrowing ($a_0 \to 0$) of the potential
 barrier to the analytic result (\ref{eq17}). In the limit case $U_0 \to \infty $ and $a_0 \to
 0$ we obtain $a_{2D}\rightarrow a_0$ for the scattering length $a_{2D}$ extracted from the calculated
 amplitude $f(q)$ by the formula (\ref{eq17}), what is in agreement to the
 estimate given in
 \cite{ref22b}. The range of applicability of Eq. (\ref{eq17}) was investigated recently in \cite{ref22h}.

\begin{table*}[htb]%[hbtp]
\caption{\label{tab1}The dependence of the scattering amplitude
$f(q,\phi,\phi_q)$ on the height of the potential barrier
(\ref{eq16}) with circular base. Calculations were carried out for
$a_0 =0.01$ with the parameters:
 $M=10^2$, $N=10^5$ and $\rho_N=15$.}
\begin{ruledtabular}
\begin{tabular}{ccccc}
&\multicolumn{2}{c}{$q=0.125$}&\multicolumn{2}{c}{$q=1$}\\
\textrm{$U_{0}$} & \textrm{$f(q,0,0)$} & \textrm{$f(q,\pi,0)$} &
\textrm{$f(q,0,0)$} & \textrm{$f(q,\pi,0)$}\\ \hline

10$^{4}$&
-0.42692 + i0.08772&
-0.42693 + i0.08772&
-0.19832 + i0.05304&
-0.19837 + i0.05304 \\
\hline
 10$^{5}$&
-0.47960 + i0.11141&
-0.47961 + i0.11141&
 -0.22999 + i0.07362&
 -0.23016 + i0.07362\\
\hline
 10$^{6}$&
-0.49055 + i0.11674&
-0.49056 + i0.11674&
 -0.23673 + i0.07853&
 -0.23696 + i0.07853\\
\hline
 10$^{8}$&
-0.49400 + i0.11846&
-0.49402 + i0.11846&
 -0.23895 + i0.07988&
 -0.23920 + i0.07988\\
 \hline
10$^{10}$&
-0.49408 + i0.11850&
-0.49409 + i0.11849 &
 -0.23899 + i0.07992&
 -0.23925 + i0.07992\\

 \hline
 \hline
Eq.(\ref{eq17})&
-0.49486 + i0.11430 &
-0.49486 + i0.11430 &
-0.23901 + i0.07952&
-0.23901 + i0.07952\\
\end{tabular}
\end{ruledtabular}
\end{table*}

\begin{figure}[hbtp]
 \centerline{\includegraphics[width=\linewidth]{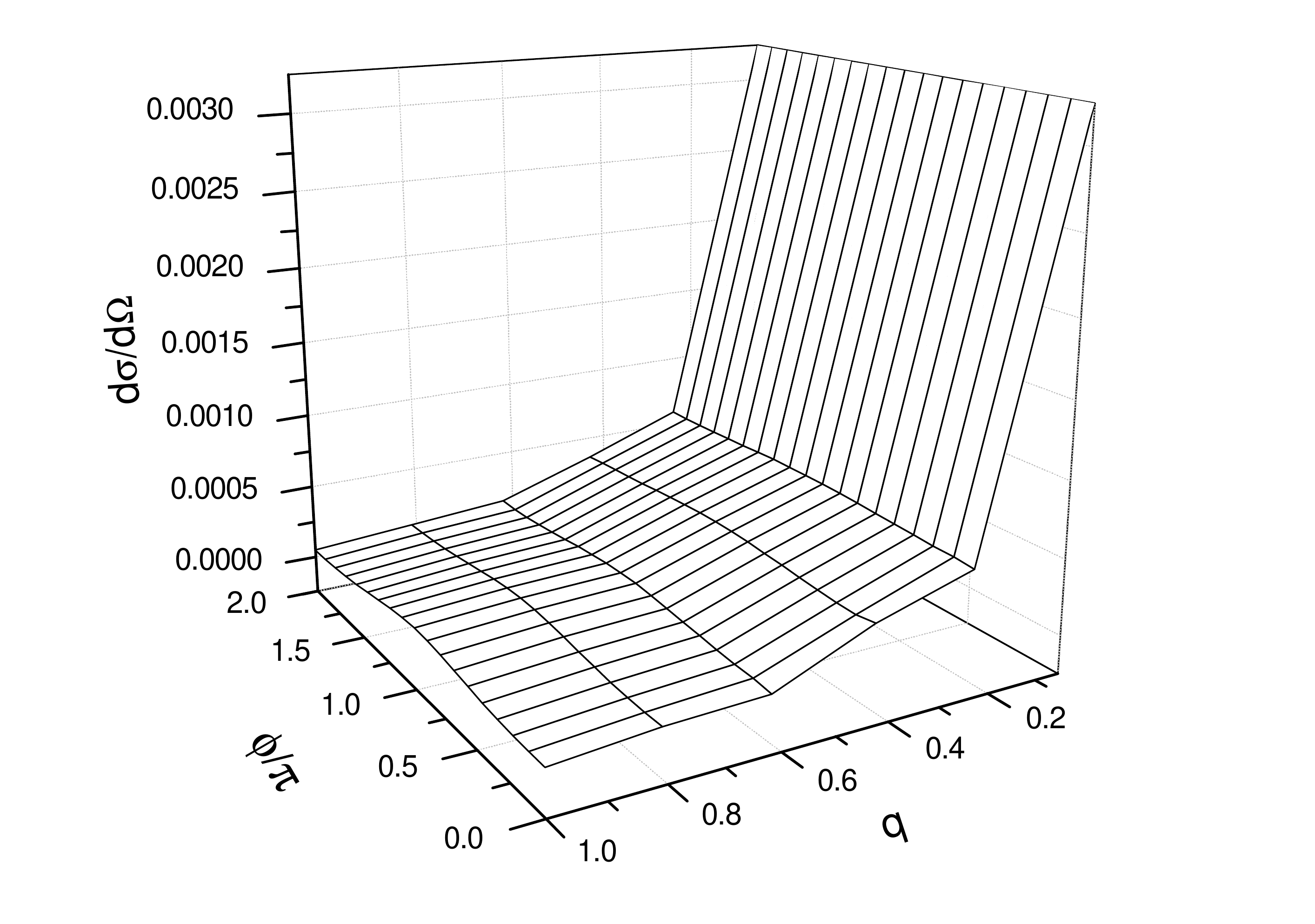}}
 \caption{The dependence of the differential cross section on the
 momentum $q$ and scattering angle $\phi$ in the case of potential barrier (\ref{eq16}) with $U_0 =10^3$
and circular base $a(\phi )=a_0=1$. Calculations were carried out
for $\phi_q=0$ with the parameters:
 $M=10^2$, $N=10^5$ and $\rho_N=15$. The quantities $d\sigma/d\Omega$ and $q$ are given in the units $\hbar=\mu=1$.}
 \label{fig9}
 \end{figure}

\begin{figure}[hbtp]
 \centerline{\includegraphics[width=\linewidth]{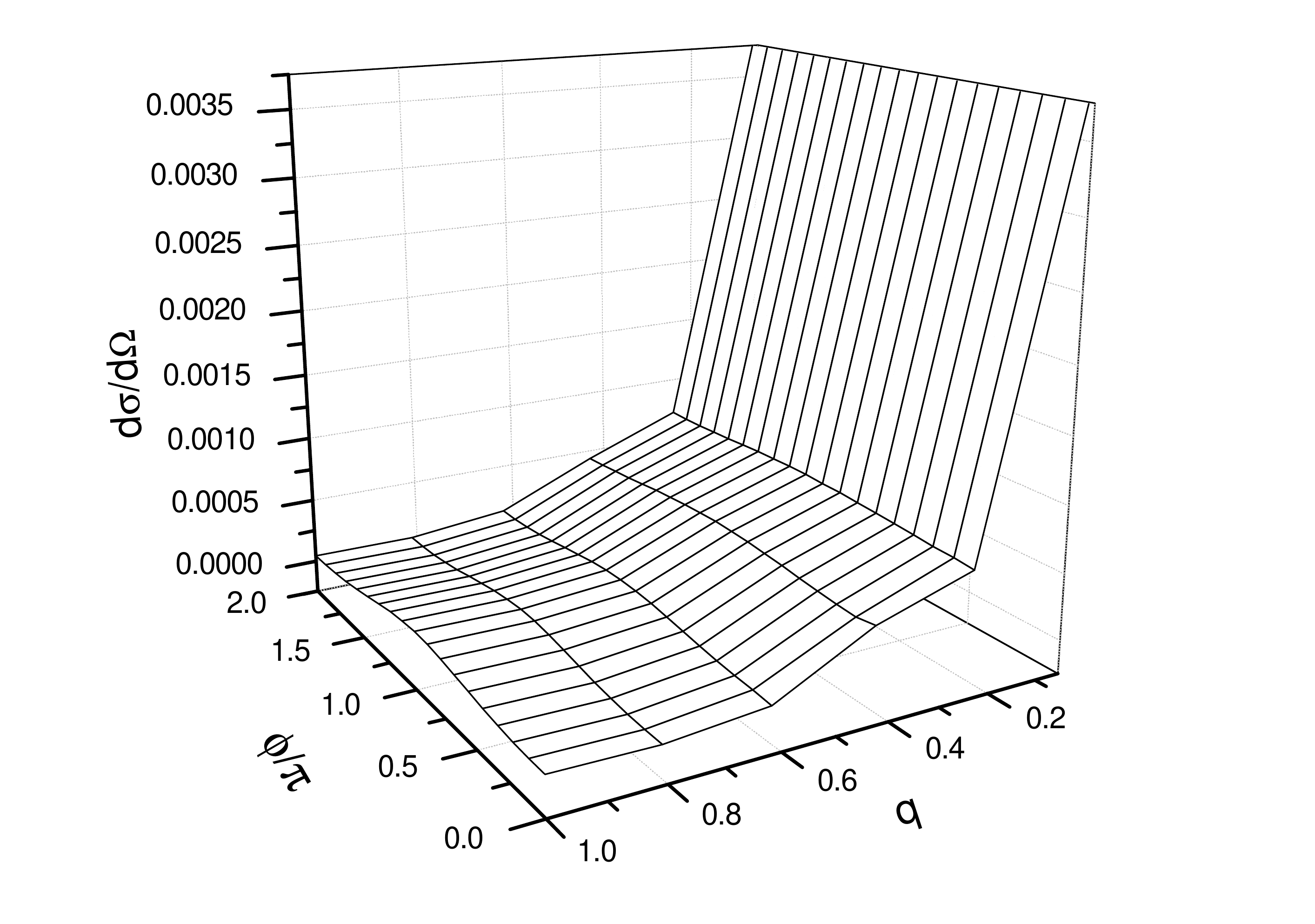}}
 \caption{The dependence of the differential cross section on the momentum
 $q$ and scattering angle $\phi$ in the case of potential barrier (\ref{eq16}) with $U_0 =10^3$ and elliptic base
 with $a(\pi /2)=1.1$ and $a(0)=1$. Calculations were carried
out for $\phi_q=0$ with the parameters:
 $M=10^2$, $N=10^5$ and $\rho_N=15$. The quantities $d\sigma/d\Omega$ and $q$ are given in the units $\hbar=\mu=1$.}
 \label{fig10}
 \end{figure}

\begin{figure}[hbtp]
 \centerline{\includegraphics[width=\linewidth]{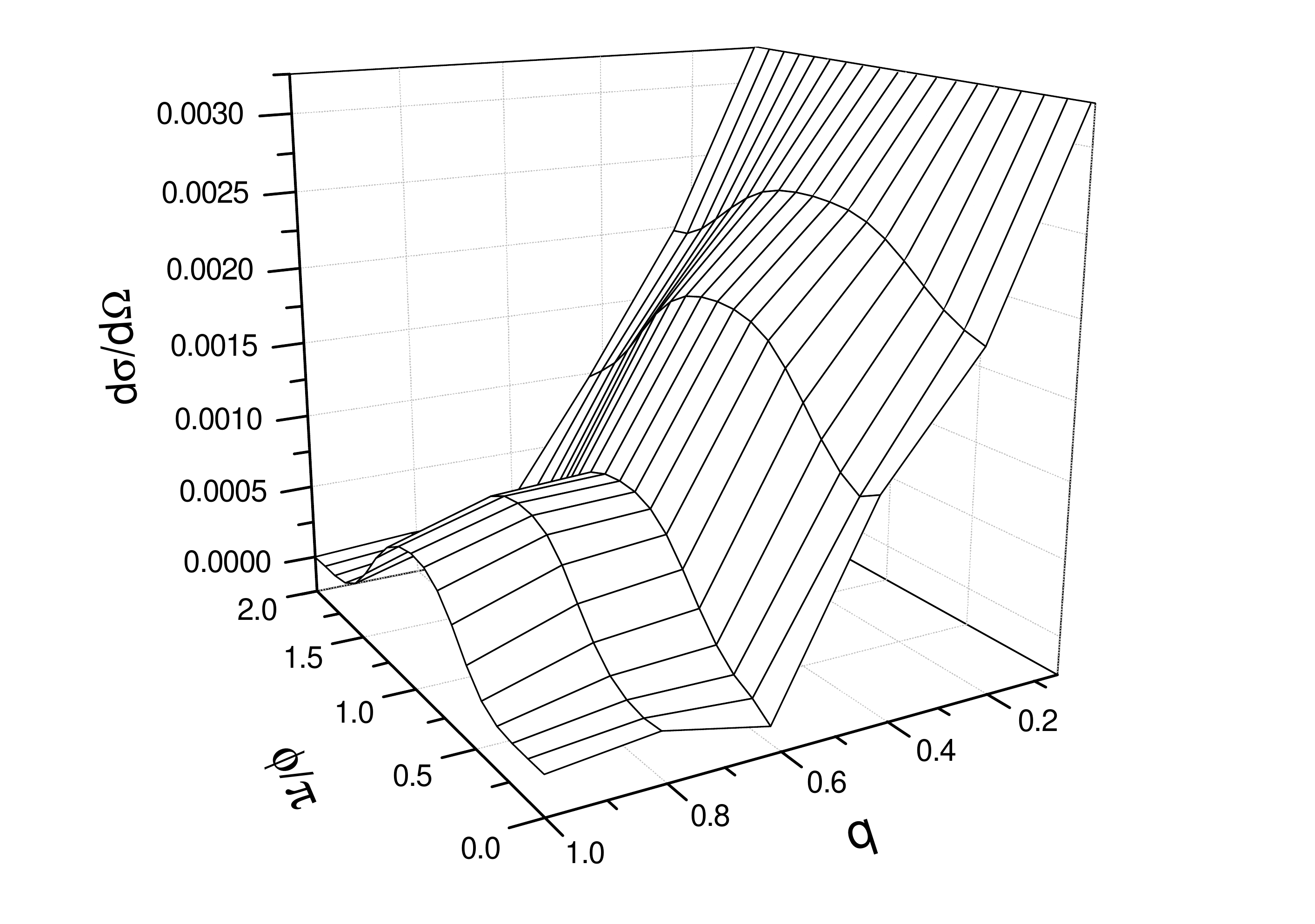}}
 \caption{The same as in Fig.~\ref{fig10} but for the case of strong anisotropy of the scatterer (\ref{eq16})
 with $a(\pi /2)=2$ and $a(0)=1$. The quantities $d\sigma/d\Omega$ and $q$ are given in the units $\hbar=\mu=1$.}
 \label{fig11}
 \end{figure}

  Then, we have applied our scheme for calculation of the scattering cross
 section ${d\sigma (q,\phi,\phi_q )}/{d\Omega
 }\mbox{=} {\left| {f(q,\phi ,\phi
 _q )} \right|^2}$   for the isotropic $\left( {a(\phi
 )=a_0 } \right)$ and anisotropic $\left( {{a(\phi =\pi
 \mathord{\left/ {\vphantom {\pi 2}} \right. \kern-\nulldelimiterspace} 2)}
 \mathord{\left/ {\vphantom {{a(\phi =\pi \mathord{\left/
 {\vphantom {\pi 2}} \right. \kern-\nulldelimiterspace} 2)} {\rho _0 (\phi
 =0)}}} \right. \kern-\nulldelimiterspace} {a(\phi =0)}=1.1 \text{ and } 2}\right)$ scattering.
In Fig.~\ref{fig9} the differential cross section, calculated for
the circular base $a(\phi)=a_0$ of the scatter (\ref{eq16}), is
given as a function of $q$ and $\phi$. The dependence of the cross
section on $\phi$  disappears with decreasing momentum $q$ and the
dependence on $\phi_q$ is absent for any $q$ due to the spherical
symmetry of the potential (\ref{eq16}) if $a(\phi)=a_0$. Further,
the analysis was extended to more general case of elliptical base
of the potential barrier (\ref{eq16}). In Figs.~\ref{fig10} and
\ref{fig11} the calculated differential cross sections on the
anisotropic scatter (\ref{eq16}) are presented for the cases of
weak ($a(\phi=\pi/2)/a(\phi=0)=1.1$) and strong anisotropy
($a(\phi=\pi/2)/a(\phi=0)=2$). Here, we observe more sharp
dependence on $\phi$ and $q$ in the cross section with increasing
anisotropy of the scatterer. The anisotropy in the scattering
cross sections appears with increasing $q$ earlier for the
anisotropic potential barrier than for the barrier with circular
base.

%
 %
 %\begin{figure}[htbp]
 %\centerline{\includegraphics[width=8.76in,height=6.68in]{Pure1.pdf}}
 %\caption{a) b)}
 %\label{fig1}
 %\end{figure}
 %\begin{figure}[htbp]
 %\centerline{\includegraphics[width=8.68in,height=6.66in]{Pure2.pdf}}
 %\caption{The dependence of a) real and b) imaginary part of the scattering amplitude
 %on the relative momentum $q$ at $\rho _0 =1$.}
 %\label{fig2}
 %\end{figure}
%

 \subsection{Dipole-dipole scattering in plane}
 Here we analyze the 2D quantum scattering on the long-range anisotropic scatterer defined by the dipole-dipole interaction.
 This problem simulates the
collisions of polar molecules in pancake optical traps. The
interaction potential between two arbitrarily
 oriented dipoles reads
 \begin{equation}
 \label{eq18}
 U({\bm \rho}, {\bm d}_1, {\bm d}_2) =\frac{1}{\rho
^3}\left(({\bm d}_1 {\bm d}_2 )-3\frac{({\bm d}_1 {\bm \rho
})({\bm d}_2 {\bm \rho })}{\rho ^2}\right)\,,
 \end{equation}
 where ${\bm d}_i ,(i=1,2)$ -- dipole moments and $({\bm d}_i {\bm \rho })/\rho$ -- their projections onto the collision axis.
 The expression (\ref{eq18}) can be written in the polar coordinates
 \begin{align}
 \label{eq19}
  U\left( {\rho ,\phi; \alpha ,\beta ,\gamma } \right)= \frac{d_1 d_2
 }{\rho ^3}[\sin (\alpha )\sin (\gamma )\cos (\beta ) + \nonumber \\
  +\cos (\alpha )\cos
 (\gamma )
  -3\sin (\alpha )\sin (\gamma )\cos (\phi )  \cos (\phi - \beta)]\,,
 \end{align}

% \begin{align}
% \label{eq19}
%  V\left( {\rho ,\phi ,\alpha ,\beta ,\gamma } \right)= \frac{d_1 d_2
% }{\rho ^3}[\sin (\alpha )\sin (\gamma )\cos (\beta )\nonumber \\
% \qquad {}  +\cos (\alpha )\cos
% (\gamma )
%  -3\sin (\alpha )\cos (\phi ) \nonumber \\
% \times \left( \sin (\gamma )\cos (\beta)\cos (\phi ) +\sin (\gamma )\sin (\beta )\sin (\phi ) \right)],
% \end{align}

\begin{figure}[hbtp]
 \centerline{\includegraphics[width=\linewidth]{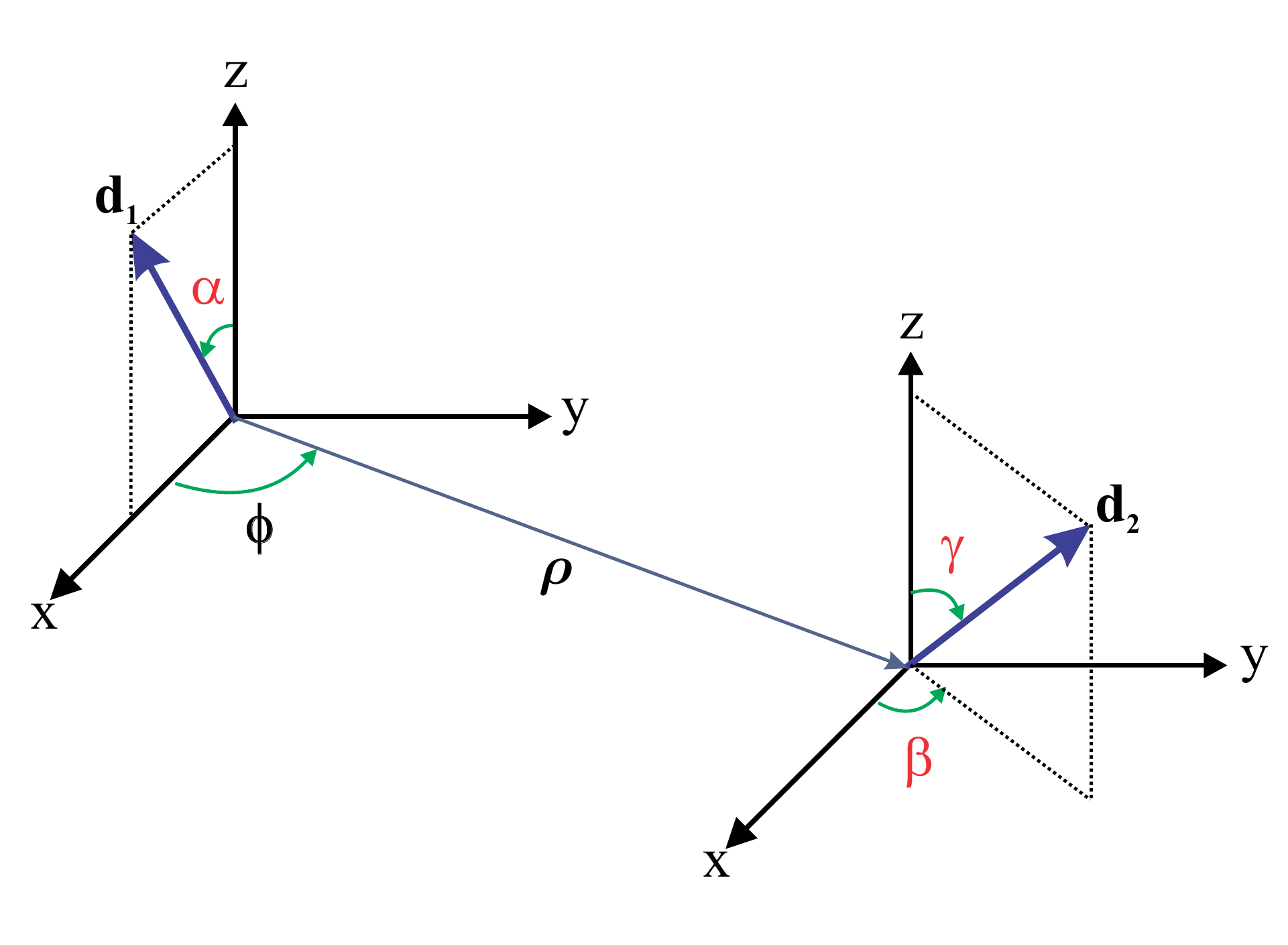}}
 \caption{(Color online) Collision in the plane $XY$ of two arbitrarily oriented dipoles ${\bm d}_1 $ and ${\bm d}_2 $}
 \label{fig12}
 \end{figure}

% \begin{equation}
% \label{eq19}
% \begin{array}{l}
%  V\left( {\rho ,\phi ,\alpha ,\beta ,\gamma } \right)=\frac{d_1 d_2
% }{\rho ^3}[\sin (\alpha )\sin (\gamma )\cos (\beta )+\cos (\alpha )\cos
% (\gamma )- \\
%  \mbox{ }-3\sin (\alpha )\cos (\phi )\left( {\sin (\gamma )\cos (\beta
% )\cos (\phi )+\sin (\gamma )\sin (\beta )\sin (\phi )} \right)], \\
%  \end{array}
% \end{equation}
 \noindent where the angles $\alpha$ and  $\gamma$ define the tilt of dipoles to the scattering plane $XY$ and the angle $\beta$
 denotes the mutual orientation of the dipole polarization planes
 $Zd_1$ and $Zd_2$ in Fig.~\ref{fig12}.

 If we consider the scenario when the polarization of colliding molecules is
 orthogonal to the plane of motion $\left( {\alpha =\beta =\gamma =0}
 \right)$, interaction is fully isotropic and repulsive
 \begin{equation}
 \label{eq20}
 U\left( \rho \right)=\frac{d_1 d_2 }{\rho ^3}\,.
 \end{equation}
 This case was intensively studied in the previous works \cite{ref16,ref18}.

\begin{figure}[hbtp]
 \centerline{\includegraphics[width=\linewidth]{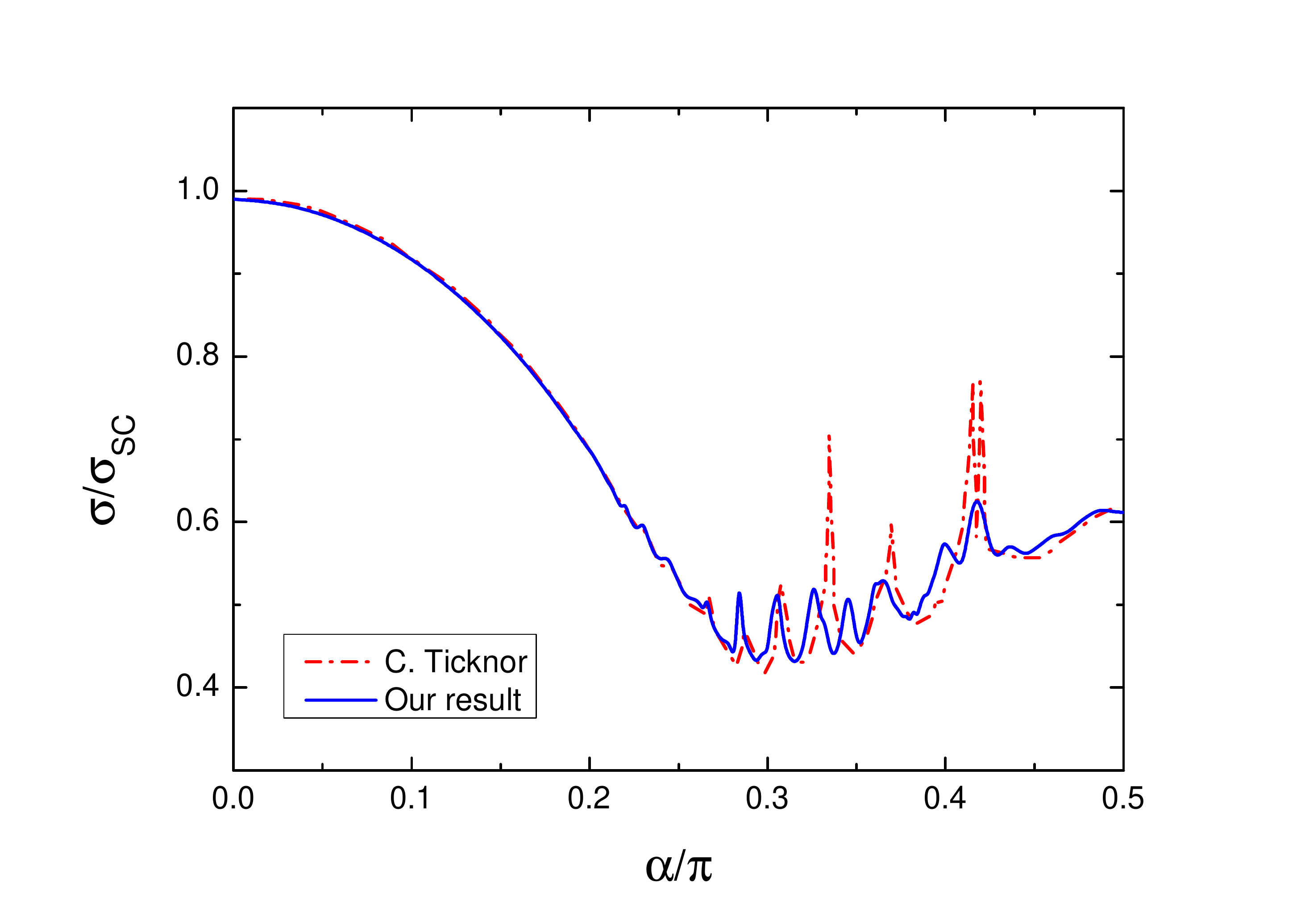}}
 \caption{(Color online) A comparison of the total cross section (in the units of $\sigma
 _{SC} $) with the result of C. Ticknor \cite{ref17} calculated for  potential (\ref{eq22}) at $D=1\,\,,\,Dq=10$.}
 \label{fig13}
 \end{figure}

 \noindent For dipoles oriented in the plane $\left( {\alpha =\gamma =\tfrac{\pi }{2}}
 \right)$, anisotropy arises and the interaction potential reads
 \begin{equation}
 \label{eq21}
 U\left( {\rho ,\phi ,\beta } \right)=\frac{d_1 d_2 }{\rho ^3}[\cos (\beta
 )-3\cos (\phi )\cos (\phi -\beta )]\,.
 \end{equation}
 A particular case of parallel dipoles with the polarization axis tilted to the
 plane of motion ($\alpha =\gamma ;\mbox{ }\beta =0)$ with short-range
 interaction modeled by a hard wall at the origin
\begin{equation}
\label{neweq1}
V_{HW}(\rho) = \left\{ {{\begin{array}{*{20}c}
  {\infty ,\mbox{ }\rho \leqslant \rho _{HW}} \hfill \\
  {0,\mbox{ }\rho >\rho _{HW}} \hfill \\
 \end{array} }} \right.
\end{equation}
  with the width ${\rho _{HW} }
 \mathord{\left/ {\vphantom {{\rho _{HW} } D}} \right.
 \kern-\nulldelimiterspace} D=0.1$
 \begin{equation}
 \label{eq22}
 U\left( {\rho ,\phi ,\alpha } \right)=V_{HW} (\rho )+\frac{d^2}{\rho
 ^3}[1-3\sin ^2(\alpha )\cos ^2(\phi )]
 \end{equation}
 was considered in paper \cite{ref17}. We have investigated this case with
 our approach and have obtained good agreement with the results of
paper \cite{ref17}.
 This is illustrated by Fig.~\ref{fig13}, where the calculated total cross
 section $\sigma(q,\alpha)$ (\ref{eq5}) is given in the units of $\sigma_{SC}$. Here $D$ is the
 dipolar length $D={\mu d^2} \mathord{\left/ {\vphantom {{\mu d^2} {\hbar
 ^2}}} \right. \kern-\nulldelimiterspace} {\hbar ^2}\mbox{ (}d=d_1 =d_2 )$
 and $\sigma _{SC} =\tfrac{4}{q}\sqrt {\pi Dq} $ is the value of the total
 scattering cross section in the eikonal approximation that is valid in the
 high-energy regime, $Dq\gg 1$ \cite{ref17}. All calculations in this section were performed for the following parameters: $M=40\,, \,N=1.2\times 10^5$ and $\rho_N =60$; the number of
 grid points on $\phi_q $ was $2M_{q}+1 = 101$.

\begin{figure}[hbtp]
 \centerline{\includegraphics[width=\linewidth]{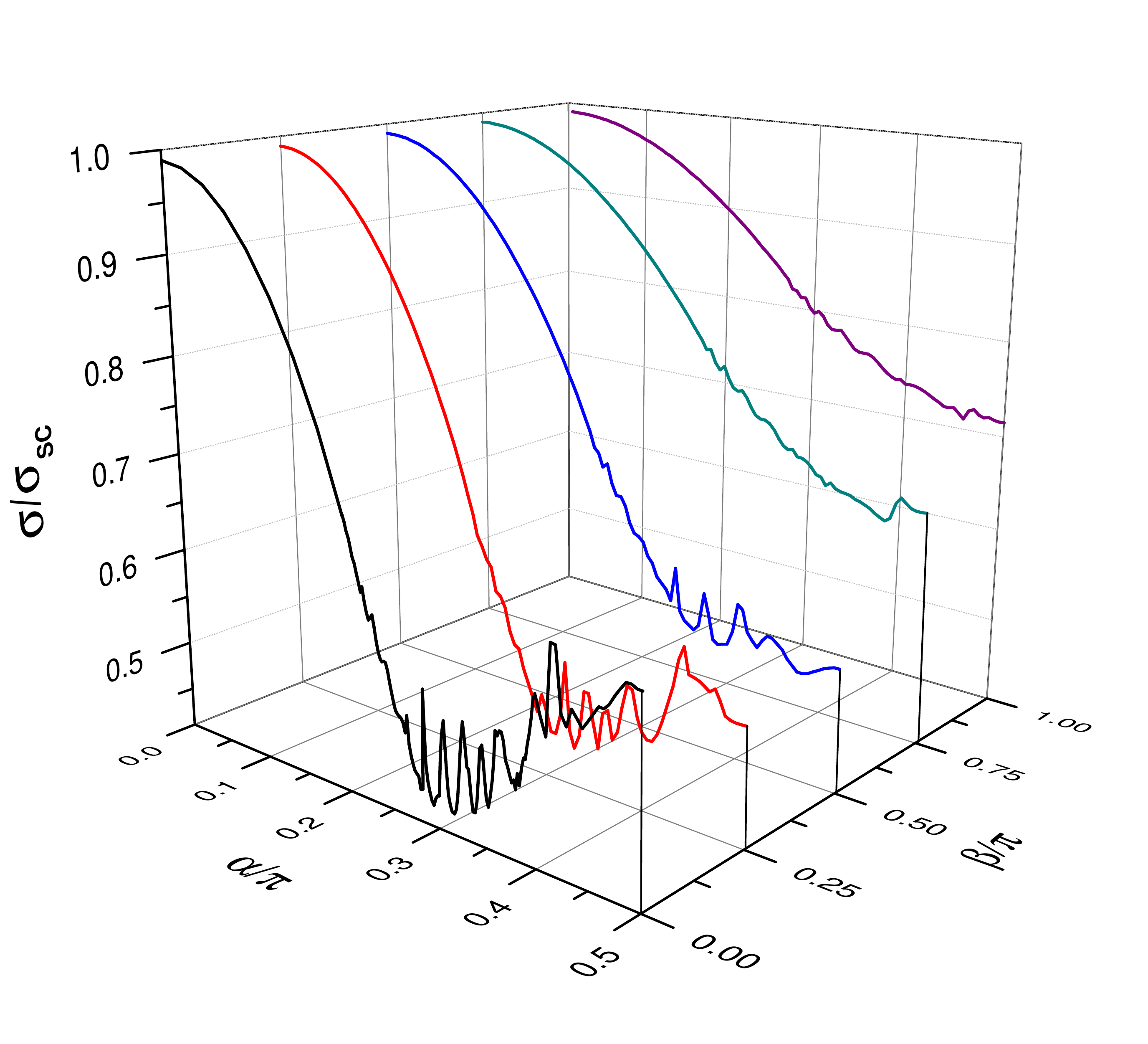}}
 \caption{(Color online)
 The total cross sections $\sigma$ in the units of $\sigma _{SC}$
  as a function of the dipole tilt angle $\alpha=\gamma$ and the rotational angle $\beta$ calculated for potential (\ref{eq19}) at $D=1\,\,,\,Dq=10$.}
 \label{fig15}
 \end{figure}

\begin{figure*}[hbtp]
\begin{minipage}[h]{0.3\linewidth}
\center{\includegraphics[height=5cm]{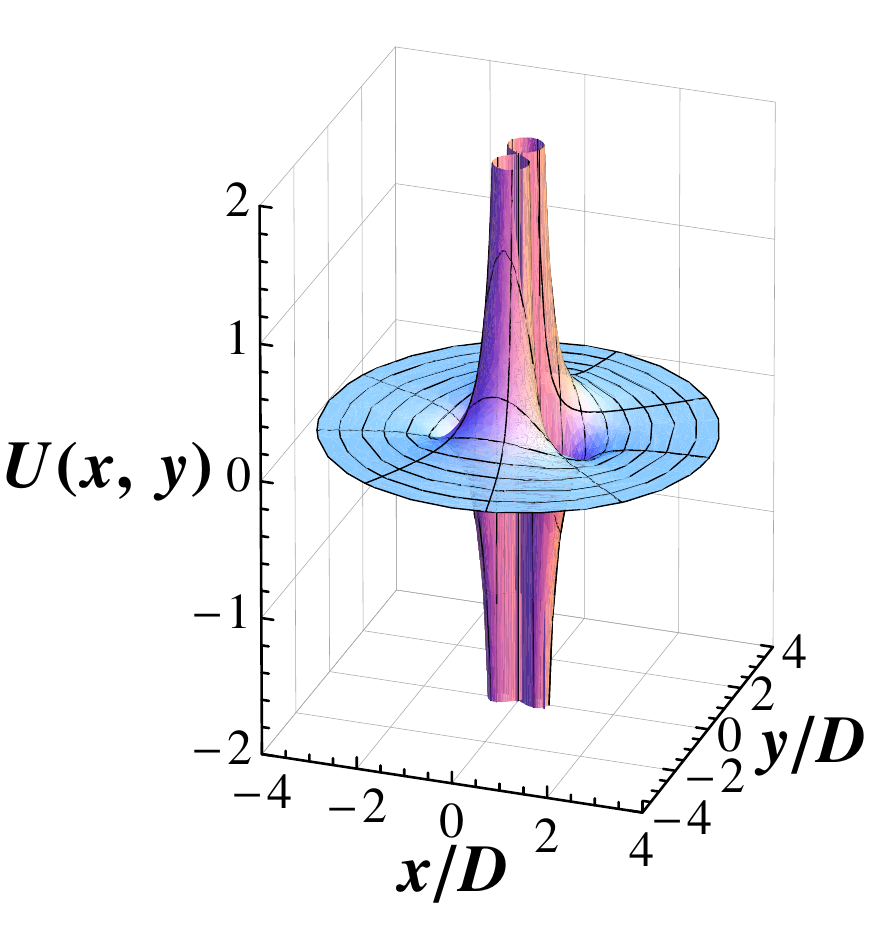}} $
\beta=0, \mbox{  } \alpha=0.25\pi$ \\
\end{minipage}
\hfill
\begin{minipage}[h]{0.3\linewidth}
\center{\includegraphics[height=5cm]{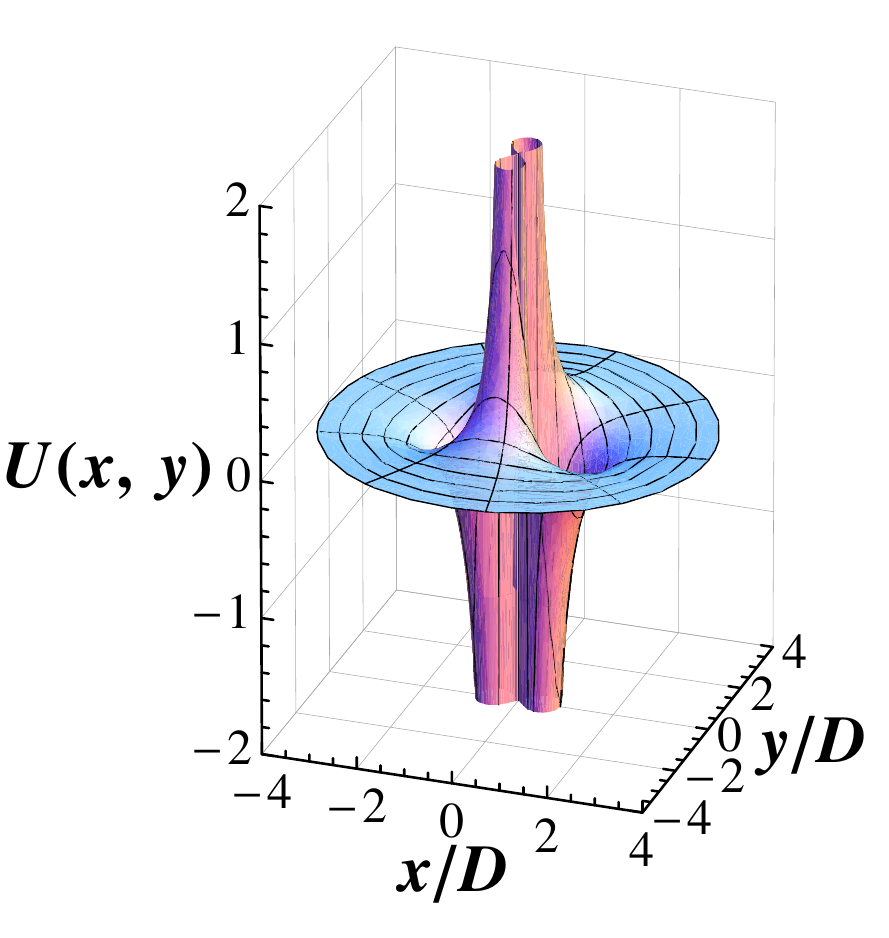}} $\beta=0, \mbox{  } \alpha=0.35\pi$ \\
\end{minipage}
\hfill
\begin{minipage}[h]{0.3\linewidth}
\center{\includegraphics[height=5cm]{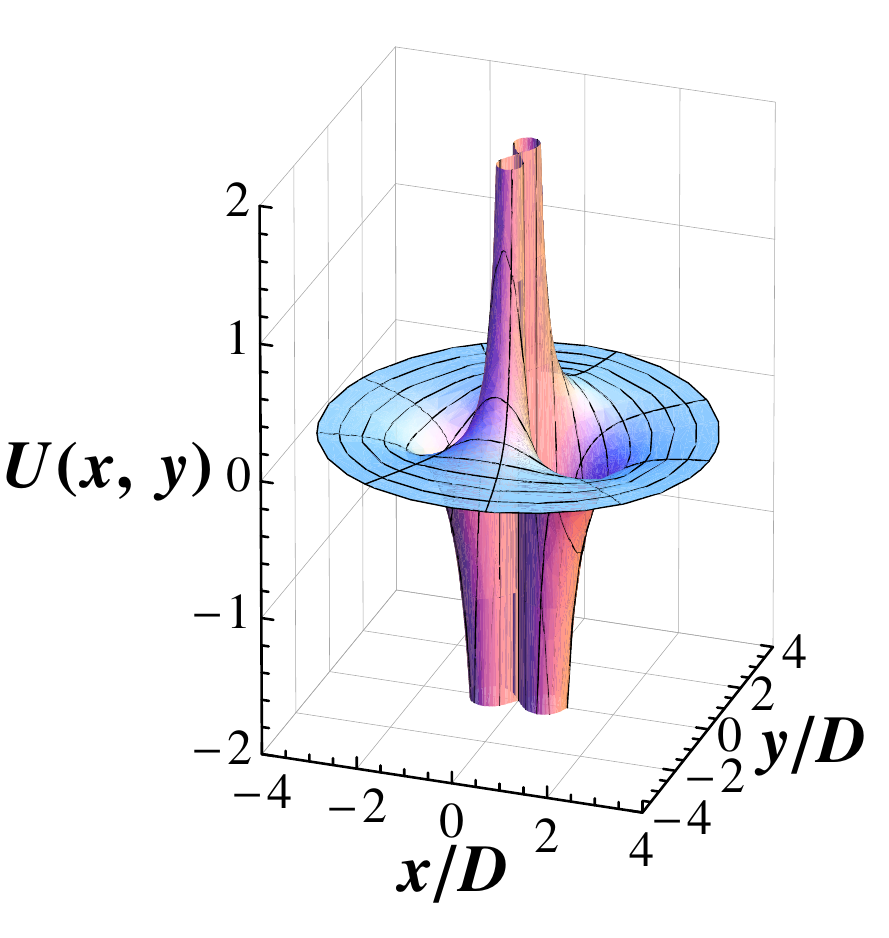}} $\beta=0, \mbox{  } \alpha=0.5\pi$ \\
\end{minipage}
\vfill
\begin{minipage}[h]{0.3\linewidth}
\center{\includegraphics[height=5cm]{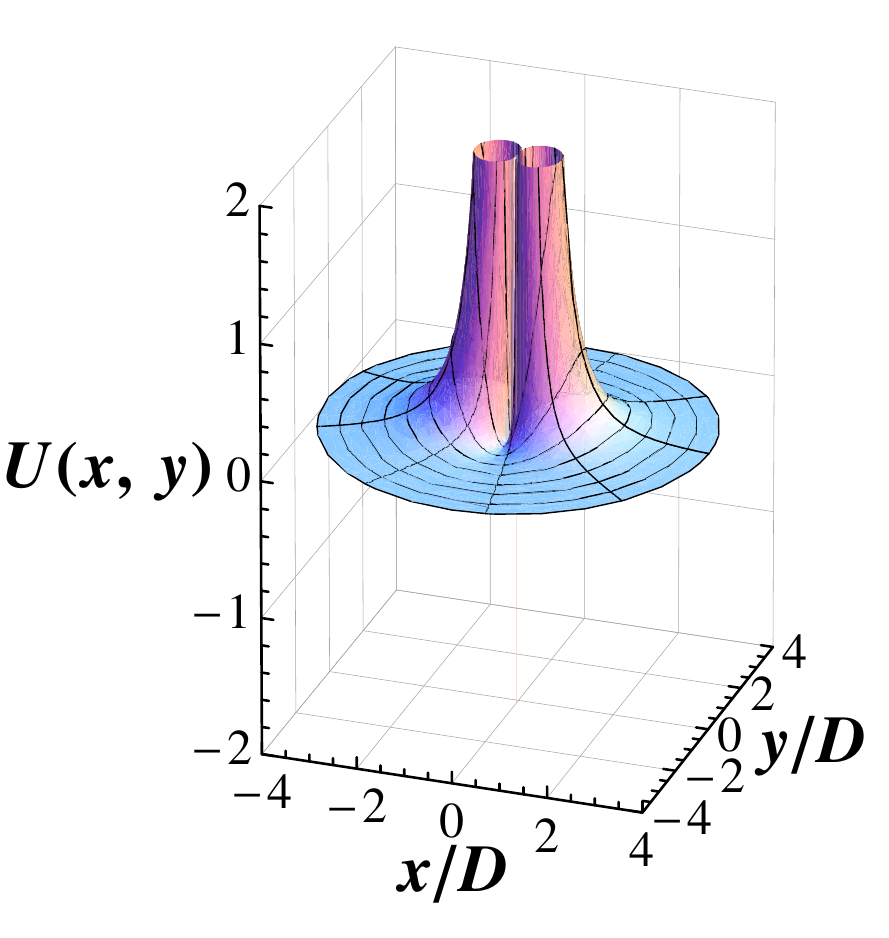}} $\beta=\pi, \mbox{  } \alpha=0.25\pi$ \\
\end{minipage}
\hfill
\begin{minipage}[h]{0.3\linewidth}
\center{\includegraphics[height=5cm]{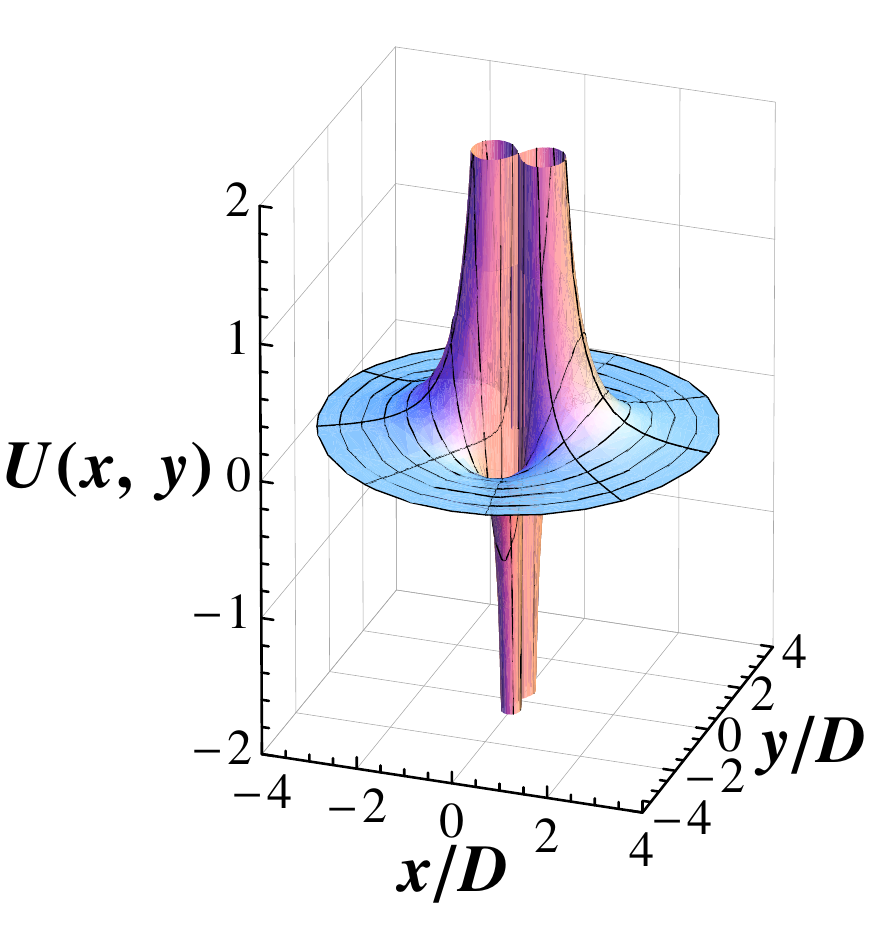}} $\beta=\pi, \mbox{  } \alpha=0.35\pi$ \\
\end{minipage}
\hfill
\begin{minipage}[h]{0.3\linewidth}
\center{\includegraphics[height=5cm]{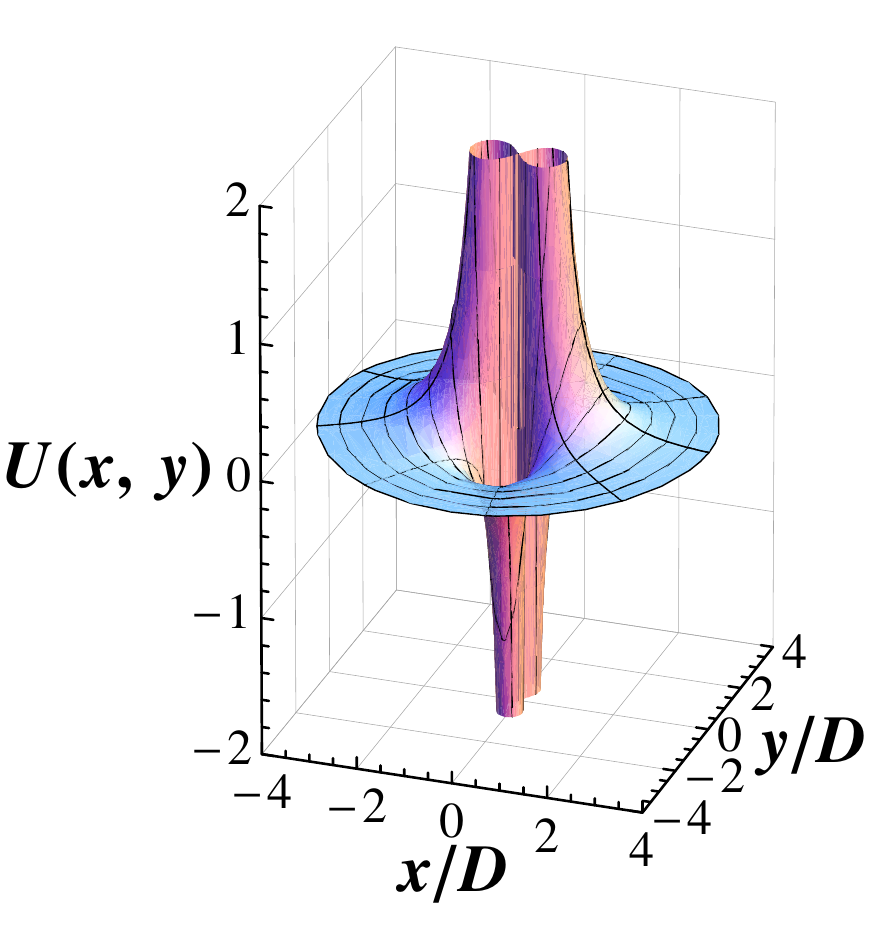}} $\beta=\pi, \mbox{  } \alpha=0.5\pi$ \\
\end{minipage}
 \caption{(Color online) The dependence of the dipole-dipole potential $U({\bm \rho},{\bm d_1},{\bm d_2})$ (\ref{eq19}) in the units
 of $E_D=\hbar^6/\mu^3 d^4$ ($d=d_1=d_2$) on the tilt angle $\alpha$ for two extreme mutual orientations $\beta=0$ and $\pi$ of
 the dipole polarization planes  $Zd_1$ and $Zd_2$.}
 \label{fig14a}
 \end{figure*}

 \begin{figure}[hbtp]
 \centerline{\includegraphics[width=\linewidth]{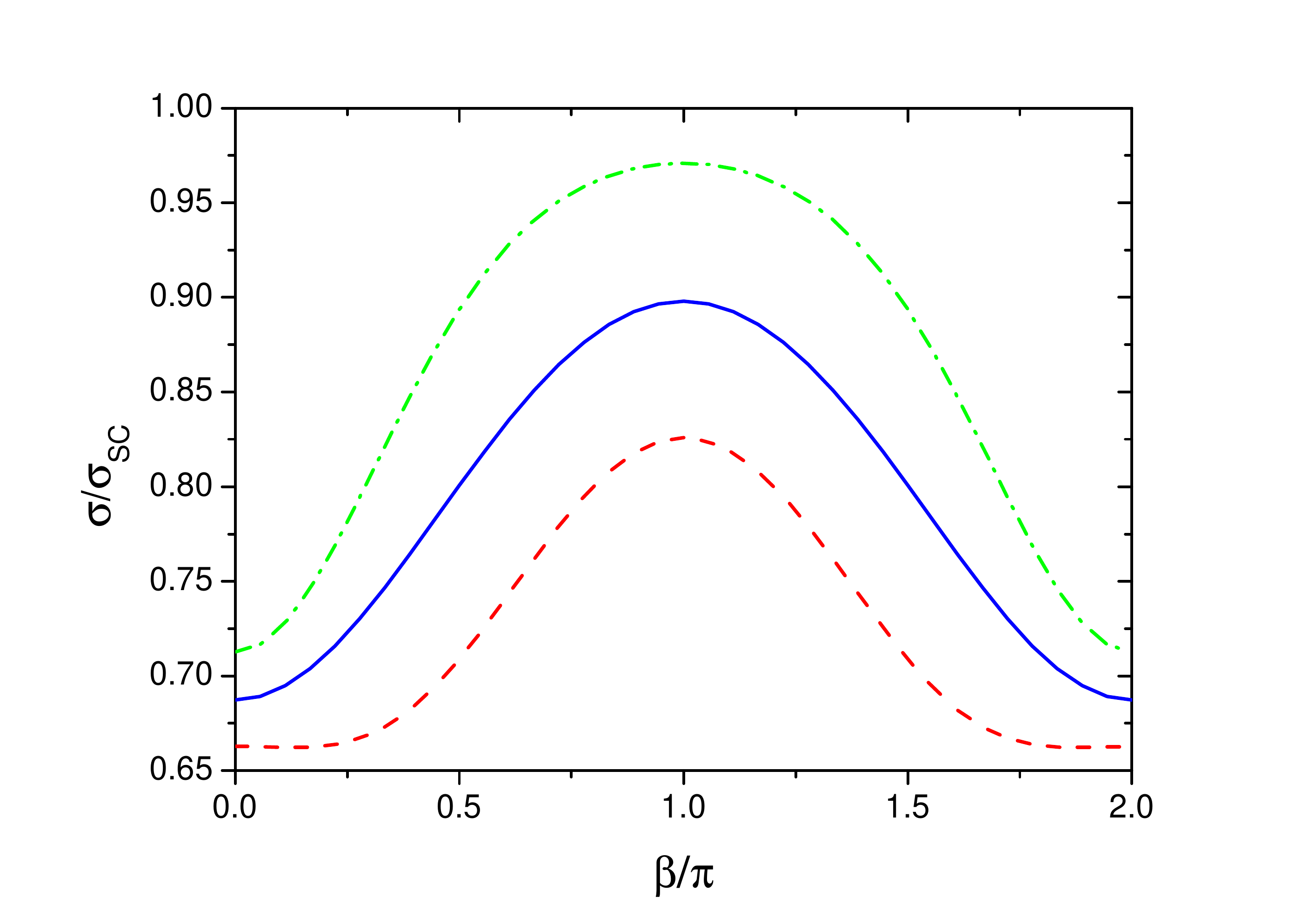}}
 \caption{(Color online)
 The total cross section $\sigma$ (solid line), bosonic cross section $\sigma_{g}$ (dashed line) and fermionic cross section $\sigma_{u}$ (dash-dot line)
 calculated for potential (\ref{eq19}) as a function of the angle $\beta$ for the fixed $\alpha=0.2\pi$ at $D=1\,\,,\,Dq=10$ (in the units of
  $\sigma _{SC}$).}
 \label{fig14}
 \end{figure}

Then, we have analyzed how the found ``resonant'' structure for
the polarized dipoles (see Fig.~\ref{fig13}) in the calculated
dependence of the scattering cross section on the dipole tilt
angle $\alpha=\gamma$ varies with destroying the polarization.
Depolarization was simulated by rotating the angle $\beta$ between
the dipole polarization planes $Zd_1$ and $ Zd_2$ (see
Fig.~\ref{fig12}). We found progressive narrowing of the
``resonance'' area with a simultaneous decrease of the amplitudes
of the ``resonance'' oscillations with increasing angle $\beta$
from $0$ to $\pi$ (see Fig.~\ref{fig15}). When approaching the
point $\pi$ the ``resonant'' structure disappears, the cross
section becomes smooth relative to $\alpha$ and reaches its
maximum value. This effect is due to the fact that when
approaching the angle $\beta=\pi$ repulsive feature of the
dipole-dipole interaction becomes dominant (see
Fig.~\ref{fig14a}). With decreasing $\beta$ from $\pi$ to $0$ the
attractive part $U(\rho,\phi)<0$ appears for some $\rho$ and
$\phi$. It leads to appearing the ``resonant" part in the
scattering cross section. Note, that the presented cross sections
were obtained for distinguishable particles. The effect of
symmetrization/antisymmetrization $f_{g,u}(\phi)=(f(\phi)\pm
f(\phi-\pi))/\sqrt{2}$ (i.e. transition to identical particles) is
shown in Fig.~\ref{fig14} for $\alpha=0.2\pi$ and $\beta$ varying
from $0$ to $2\pi$, where we
 observe the strong dependence of the total cross sections on the angle $\beta$
 with the maximal enhancement at $\beta =\pi$. The cross sections are symmetric with respect to this
 point. We have to note that the curves $\sigma_g(\beta)$ and
 $\sigma_u(\beta)$ describing the scattering of bosonic and
 fermionic particles exactly repeat the behavior of the curve
 $\sigma(\beta)$ for distinguishable particles.

\begin{figure}[hbtp]
 \centerline{\includegraphics[width=\linewidth]{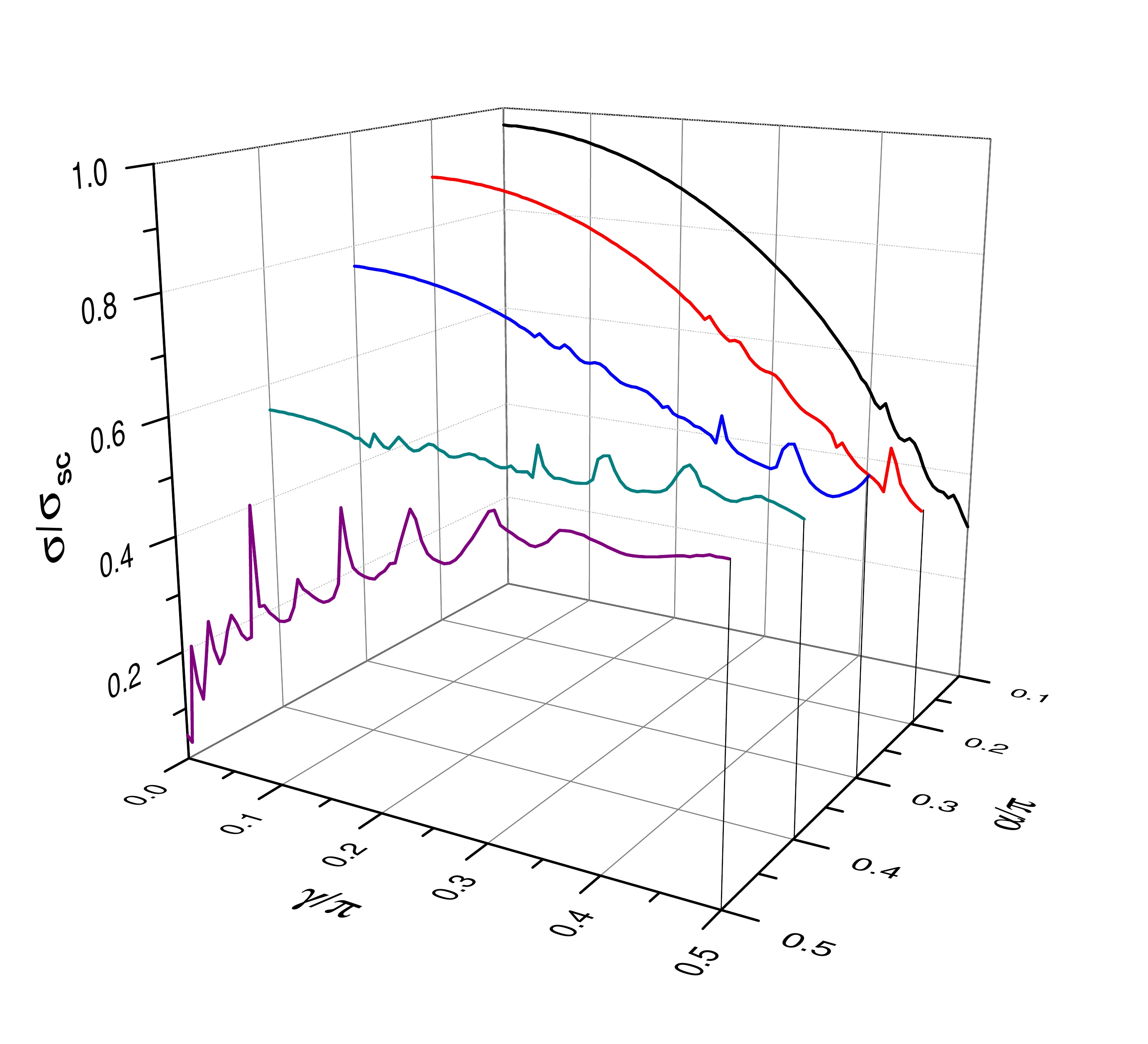}}
 \caption{(Color online)
 The total cross sections $\sigma$ in the units of $\sigma _{SC}$
  as a function of the dipole tilt angles $\alpha$ and $\gamma$ calculated for potential (\ref{eq19}) at $D=1\,\,,\,Dq=10$.
   The rotational angle $\beta$ is equal to $\pi/2$.}
 \label{fig16}
 \end{figure}

Finally, we have analyzed the scattering of arbitrarily oriented
dipoles in the case of mutual orthogonality of their polarization
planes $ Zd_1$ and $Zd_2$ ($\beta = \pi /2$). Here also we found a
strong ``resonant'' structure by the tilt angle $\gamma$ of one
dipole, if the other dipole is oriented in the scattering plane
$XY$  ($\alpha=\beta=\pi/2$) (see Fig.~\ref{fig16}), that appears
due to the attractive feature of the dipole-dipole interaction
strength with increasing of the tilt angle $\alpha\rightarrow
\pi/2$.

\subsection{Advantages and prospects of the angular-grid representation}
Let us discuss here the advantages of our angular-grid
representation (\ref{eq7}) in comparison with the traditional
partial-wave formalism, which was used in particular by Ticknor in
the analysis of the 2D scattering of polarized dipoles
\cite{ref17}.

First, for any potential $U(\rho,\phi)$ the estimate of the
residual term $\delta\Psi_M$ in the expansion
$\Psi(\rho,\phi)=\sum_{m=-M}^{M}\chi_m(\rho)e^{im\phi}/\sqrt{\rho}+\delta\Psi_M$
of the desired wave function on partial waves is unknown. This
fact becomes crucial in the case of strongly anisotropic and
long-range potential considered in the above Subsection III.B,
where the partial-wave expansion becomes very slow due to the
strong coupling between different angular momenta remaining even
at the zero-energy limit \cite{ref14,ref15}. Opposite, our
angular-grid representation belongs to a class of mesh methods
admitting an estimate of an approximation error. Thus, since the
representation (\ref{eq7}) can be considered, following
\cite{ref21}, as a Fourier interpolation of the order $M$ in the
variable $\phi$, the error of the approximation (\ref{eq7}) can be
estimated as \cite{ref29}
$$
\delta\Psi_M=|\Psi(\rho,\phi)-\sum_{j=0}^{2M}\omega_j(\phi)\Psi(\rho,\phi_j)|<
 const\frac{\ln M}{M^k}\,,
$$
where $\omega_j(\phi)=\frac{2\pi}{(2M+1)}\sum_{m=-M}^{M}
e^{im(\phi-\phi_j)}$ and $k$ is the number of existing continuous
and bounded derivatives over $\phi$. Due to this estimate one can
await a rapid convergence of our angular-grid representation
(\ref{eq7}) with increasing $M$ on the sequence of compressed
grids $\{\phi_j\}_0^{2M}$. It was confirmed in all the
computations we performed (see also illustration of the
convergence of the angular-grid representation given in Table.II
of Appendix B). The convergent results were obtained even in the
regions of ``resonant'' scattering of polarized as well as
unpolarized dipoles (see Figs.7,8 and 11). Slight shifts of the
positions of the ``resonances'' in the scattering of polarized
dipoles calculated with our method relative to the values obtained
by Ticknor (see Fig.7) can be explained by the error due to the
truncation of his partial-wave sum.

Another important advantage of the angular-grid representation is
its flexibility connected with the lack of the time-consuming
procedure of calculating the matrix elements of the interaction
potential. As was mentioned above, the potential matrix in this
representation is diagonal and consists of the values of the
potential in the angular grid nodes. This circumstance permitted
us to perform extended detail computations of the 2D dipole-dipole
scattering of the different mutual orientations. This also makes
the method very perspective in the case of nonseparable
interactions and generalization to higher dimensions.

%\newpage ~
%\newpage

\section{Conclusion}
\label{sec:conclusion} We have developed a computational scheme
for quantitative analysis of the 2D quantum scattering on the
long-range anisotropic potentials. High efficiency of the method
was demonstrated in the analysis of scattering on the cylindrical
potential with the elliptical base and dipole-dipole collisions in
the plane. In the last case  we found the strong dependence of the
scattering cross section on the mutual orientation of dipoles.

The method can be applicable for analyzing the collisional
dynamics of the polarized as well as unpolarized polar molecules
in 2D and quasi-2D traps. A natural application is the
quantitative analysis of the confinement-induced resonances (CIR)
in quasi-2D traps. In this problem the crucial element is an
inclusion of the transverse confinement what we suppose to explore
in our future work with generalization of the developed approach
to this 3D case. Particularly, this analysis can resolve the
puzzle with the position of the 2D CIR measured recently
\cite{ref27}, which is under intensive discussions.

\begin{acknowledgments}
The authors thank V.~V.~Pupyshev and V.~B.~Belyaev for fruitful
discussions and comments. The authors acknowledge the support by
the Russian Foundation for Basic Research, grant 14-02-00351.
\end{acknowledgments}

\begin{table*}[hbt]%[hbt]%[hbtp]
\caption{\label{tab4} The dependence of the scattering amplitude
$f(q,\phi,\phi_q)$ on the number of angular-grid points for the
scatterer (\ref{eq16}) with elliptical base. Calculations were
performed for $q=1.5$
 with the parameters: $U_0 =10^3, \quad N=10^4$, and $\rho_N=15$.}
\begin{ruledtabular}
\begin{tabular}{ccccc}
&\multicolumn{2}{c}{$\mbox{ }a(\pi /2)/ a(0)=1.1$ }&\multicolumn{2}{c}{$\mbox{ }a(\pi /2) /a(0)=2.0$}\\
\textrm{$M$} & \textrm{$f(q,0,0)$} & \textrm{$f(q,\pi,0)$} &
\textrm{$f(q,0,0)$} & \textrm{$f(q,\pi,0)$}\\ \hline

1&
 0.94212 - i0.36446&
-0.33027 + i1.24225&
-0.95099 - i0.52976&
 0.36649 + i1.04884 \\
\hline
2&
0.65275 -- i0.30331&
-0.61946 + i1.30336&
-1.48979 + i0.79336&
-0.50755 + i2.02677\\
\hline
3&
0.68693 -- i0.30254&
-0.65392 + i1.30275&
-0.31229 + i0.79336&
-1.01013 + i1.68129 \\
\hline
4&
0.68794 -- i0.30507&
-0.65902 + i1.29149&
 -0.28629 + i0.18596&
-0.90989 + i1.50458 \\
\hline
5&
0.68799 -- i0.30506&
-0.65940 + i1.29164&
-0.50155 + i0.05523&
-0.87486 + i1.52425 \\
\hline
10&
&
&
-0.60611 + i0.07641&
-0.89428 + i1.53586 \\
\hline
20&
&
&
-0.61098 + i0.07836&
-0.89552 + i1.53816 \\
\hline
30&
&
&
-0.61173 + i0.07971&
-0.89572 + i1.53885 \\
\end{tabular}
\end{ruledtabular}
\end{table*}

\newpage

\appendix

\section{Finite-difference approximation for boundary-value problem (10),(11) and (15)}

 The boundary-value problem
(\ref{eq10}),(\ref{eq11}) and (\ref{eq15}), obtained in Section II
in the angular-grid representation (\ref{eq7}), reads in a matrix
form as:

 \begin{equation}
 \label{eq23}
 \left\{ {\begin{array}{l}
  \rho ^2\frac{d^2}{d\rho ^2}\bm {\psi}(\rho)+\left[ {\frac{1}{4}\hat{I}+ \frac{2\mu}{\hbar^2}  \rho
 ^2\left( {E \hat{I}-\hat{U}(\rho )} \right)} \right]\bm {\psi }(\rho )\,+\\
 \hspace{5cm} +\,\hat{h}^{(0)}
 \bm {\psi }(\rho )=0 \\
  \bm {\psi }\xrightarrow[{\rho \to 0}]{}const\cdot \sqrt \rho \\
  \frac{2\pi}{(2M + 1)\sqrt \rho }\sum\limits_{j=0}^{2M } {e ^{-im\phi_j} \psi _j
 (\rho )} =i^mJ_m (q\rho )\sqrt {2\pi } + \\
\hspace{5cm} + \,\frac{f_m(\phi_q)}{\sqrt {-i\rho }
 }e^{iq\rho } \,,\\
  \end{array}} \right.
 \end{equation}
 where \mbox{$h^{(0)}_{jj'}=-\frac{2\pi}{(2M + 1)}\sum\limits_{m=-M}^{M}{m^2  e^{im(\phi_j-\phi_{j'})}
 \,,\,\hat{U}(\rho)=}$} \mbox{$\mathrm{diag}\left( {U(\rho, \phi_q,\phi _0
 ),U(\rho,\phi_q,\phi _1 ),\ldots, U(\rho,\phi_q,\phi _{2M })} \right)$} and $\bm {\psi}(\rho)=\{\psi(\rho,\phi_0),\psi(\rho,\phi_1),\ldots\psi(\rho,\phi_{2M})\}$.
 In this representation the angular dependence is built into the
 matrix $\hat{h}^{(0)}$ and the interaction is included into the diagonal matrix $\hat{U}(\rho )$ of
 values of the potential in
 the angular grid nodes. The constant matrix $\hat{h}^{(0)}$ couples all equations in a system
and does not depend on the radial variable. There is no need to
compute any matrix elements of the potential, what essentially
minimizes the computational costs.

 For solving boundary value problem (\ref{eq23}) the seven-point finite-difference approximation for
 second derivatives of sixth order
 \begin{align}
 \label{eq24}
 \frac{d^2}{d\rho ^2}\bm {\psi }_{n} =\frac{1}{180h^2}
 \left(
 2\bm {\psi }_{n-3}-27\bm {\psi
 }_{n-2}+270\bm {\psi }_{n-1}\nonumber
\right.
  \\
  \left.
 -490\bm {\psi }_{n}+270\bm {\psi }_{n+1}-27\bm
 {\psi }_{n+2}+2\bm {\psi }^{n+3}\right)
  + O(h^6)
 \end{align}
%  \begin{equation}
% \label{eq24}
% \frac{d^2}{d\rho ^2}\vec {\psi }_j =\frac{2\vec {\psi }^{j-3}-27\vec {\psi
% }^{j-2}+270\vec {\psi }^{j-1}-490\vec {\psi }^j+270\vec {\psi }^{j+1}-27\vec
% {\psi }^{j+2}+2\vec {\psi }^{j+3}}{180h^2}
% \end{equation}
is applied in the points $\rho_n$ of the radial grid, where $\bm
{\psi }_{n}\equiv\bm {\psi }(\rho_{n}), \mbox{
 (}n=3,4,\ldots ,N-3\mbox{)}$. As a result, the system (\ref{eq23}) reduces to the system of linear algebraic equations (\ref{my_eq16}) with the matrix $\hat{A}$ whose band structure reads
  \begin{equation}
 \label{eq25}
 \sum_{n'=n-3}^{n+3}\hat{A}_{nn'} \bm {\psi}_{n}=\bm {F}_{n}\,\,,
 \end{equation}
%  \begin{equation}
% \label{eq25}
% A_j \vec {\psi }^{j-3}+B_j \vec {\psi }^{j-2}+C_j \vec {\psi }^{j-1}+D_j
% \vec {\psi }^j+E_j \vec {\psi }^{j+1}+F_j \vec {\psi }^{j+2}+G_j \vec {\psi
% }^{j+3}=\vec {H}_j
% \end{equation}

\noindent where the $\hat{A}_{nn'}$ coefficients are the square
$(2M+1) \times (2M+1)$ matrices, and the elements $\bm {F}_{n}$ of
the right-side of (\ref{eq25}) are the $2M+1$ dimensional vectors.
After employing the ``right-side'' boundary condition in the form
(\ref{my_eq16}) in the last three grid points $n=N-2,N-1,N$ and,
analogously, the ``left-side'' boundary condition in the form
(\ref{my_eq17}) in the first two points $n=1,2$, the detailed
structure of the matrix $\hat{A}$ is represented as

 \begin{equation}
 \label{eq26}
 \left\{ {\begin{array}{l}
 \sum_{n'=n}^{n+1} \hat{A}_{nn'} \bm {\psi }_{n'}=\bm {F}_n\,\,,\,\,\,\,\,\,n=1,2 \\
 \sum_{n'=n-3}^{n+3} \hat{A}_{nn'} \bm {\psi }_{n'}=0 \,\,,\,\,\,\,\,n=3,4,..,N-3\\
 \sum_{n'=n-1}^{n}\hat{A}_{nn'} \bm {\psi }_{n'}=\bm {F}_{n}\,\,,\,\,n=N-2,N-1,N\,. \\
  \end{array}} \right.
 \end{equation}
 The block structure of the system (\ref{eq26}) provides several significant
 advantages. The block matrix can be stored in a packaged form, which allows the use of
 optimal resource. The system (\ref{eq26}) can be efficiently solved by a fast
 implicit matrix algorithm based on the idea of the block sweep method
 \cite{ref25}.

 \section{Convergence of computational scheme}

%\begin{table*}[hbt]%[hbtp]
%\caption{\label{}}
%\begin{ruledtabular}
%\begin{tabular}{ccccc}
%&\multicolumn{2}{c}{A2}&\multicolumn{2}{c}{A1}\\
%A&B&C&D&E\\ \hline
%0&
%0.0625&
%0.00807 + i0.00051&
%0.01013 + i0.00062&
%0.01014 + i0.00063 \\
%\hline
%0&
%0.1250&
%0.00807 + i0.00047&
%0.01013 + i0.00057&
%0.01015 + i0.00057 \\
%\end{tabular}
%\end{ruledtabular}
%\end{table*}

In the Table \ref{tab4} we illustrate the convergence of the
calculated scattering amplitude $f(q,\phi,\phi_q)$ over the number
of angular grid points $2M+1$ for the
 scatterers with weak and essential anisotropy at ${a(\pi \mathord{\left/
 {\vphantom {\pi 2}} \right. \kern-\nulldelimiterspace} 2)} \mathord{\left/
 {\vphantom {{a(\pi \mathord{\left/ {\vphantom {\pi 2}} \right.
 \kern-\nulldelimiterspace} 2)} {a(0)}}} \right.
 \kern-\nulldelimiterspace} {a(0)}=1.1\mbox{ and }2$ in the potential barrier (\ref{eq16}). For the case
 ${a(\pi \mathord{\left/ {\vphantom {\pi 2}} \right.
 \kern-\nulldelimiterspace} 2)} \mathord{\left/ {\vphantom {{a(\pi
 \mathord{\left/ {\vphantom {\pi 2}} \right. \kern-\nulldelimiterspace} 2)}
 {a(0)}}} \right. \kern-\nulldelimiterspace} {a(0)}=1.1$ we
 reach the accuracy of four significant digits in the scattering
 amplitude on the angular grids with $M = 5$. For
 stronger anisotropy ${a(\pi \mathord{\left/ {\vphantom {\pi 2}}
 \right. \kern-\nulldelimiterspace} 2)} \mathord{\left/ {\vphantom {{a
 (\pi \mathord{\left/ {\vphantom {\pi 2}} \right. \kern-\nulldelimiterspace}
 2)} {a(0)}}} \right. \kern-\nulldelimiterspace} {a(0)}=2$ the
 accuracy of two significant digits was reached at $M =30$.

The number of radial grids $N$ and the border of integration
$\rho_N$ were chosen to keep the accuracy of four significant
digits in the calculated amplitudes.

%\FloatBarrier
%\clearpage

%\vspace*{\baselineskip}


\newpage

 \begin{thebibliography}{33}
 \bibitem{ref1} M.~A.~Baranov, Phys. Rep. {\bf 464}, 71 (2008).
 \bibitem{ref2} I.~Bloch, J.~Dalibard and W.~Zwerger, Rev. Mod. Phys. {\bf 80}, 885 (2008).
 \bibitem{ref3} C.~Ticknor, R.~M.~Wilson and J.~L.~Bohn, Phys. Rev. Lett. {\bf 106}, 065301 (2011).
 \bibitem{ref4} G.~M.~Brunn and E.~Taylor, Phys. Rev. Lett. {\bf 101}, 245301 (2008).
 \bibitem{ref5} J.~C.~Cremon, G.~M.~Brunn and S.~M.~Reimann, Phys. Rev. Lett. {\bf 105}, 255301 (2010).
 \bibitem{ref6} K.-K.~Ni et. al., Science {\bf 322}, 231 (2008);
 S.Ospelkaus et. al., Science {\bf 327}, 853 (2010).
 \bibitem{ref7} L.~D.~Carr et. al., New J. Phys {\bf 11}, 055049 (2009).
 \bibitem{ref8} M.~H.~G.~de~Miranda et. al., Nat. Phys. {\bf 7}, 502 (2011).
 \bibitem{ref9} P.~Minnhagen, Rev. Mod. Phys. {\bf 59}, 1001 (1987).
 \bibitem{ref10} P.~A.~Lee, N.~Nagaosa and X.-G.~Wen, Rev. Mod. Phys. {\bf 78}, 17 (2006).
 \bibitem{ref11} K.~S.~Novoselov, Rev. Mod. Phys. {\bf 83}, 837 (2011).
 \bibitem{ref12} C.~Nayak, S.~H.~Simon, A.~Stern, M.~Freedman and S.~D.~Sarma, Rev. Mod. Phys. {\bf 80}, 1083 (2008).
 \bibitem{ref13} K.~Martiyanov, V.~Makhalov and A.~Turlapov, Phys. Rev. Lett. {\bf 105}, 030404 (2010); A. Turlapov, JETP Letters {\bf 95}, 96 (2012).
 \bibitem{ref14} M.~Marinescu and L.~You, Phys. Rev. Lett. {\bf 81}, 4596 (1998); B.~Deb and L.~You. Phys. Rev. A {\bf 64}, 022717 (2001)
 \bibitem{ref15} V.~S.~Melezhik and Chi-Yu~Hu, Phys. Rev. Lett. {\bf 90}, 083201 (2003).
 \bibitem{ref16} C.~Ticknor, Phys. Rev. A {\bf 80}, 052702 (2009).
 \bibitem{ref17} C.~Ticknor, Phys. Rev. A {\bf 84}, 032702 (2011).
 \bibitem{ref18} C.~Ticknor, Phys. Rev. A {\bf 81}, 042708 (2010).
 \bibitem{ref19} J.~P.~D'Incao and C.~H.~Greene, Phys. Rev. A {\bf 83}, 030702 (2011).
 \bibitem{ref20} Z.~Li, S.~V.~Alyabishev and R.~V.~Krems, Phys. Rev. Lett. {\bf 100}, 073202 (2008).
 \bibitem{ref21} V.~S.~Melezhik, J. Comput. Phys. {\bf 92}, 67-81 (1991).
% \bibitem{ref22} L.D.~Landau, E.M.~Lifshitz, Quantum Mechanics Butterworth-Heinemann, Oxford, 1999.
%\bibitem{ref22} L.D.~Landau, E.M.~Lifshitz, Quantum mechanics, Pergamon Press, 1977.

%22
\bibitem{ref22b} L.~D.~Landau and E.~M.~Lifshitz, in \textit{Quantum mechanics: Non-Relativistic Theory}, Vol. 3 (Pergamon Press, 1977) 3rd ed., Chap. 132, pp. 551--552.


%23
 \bibitem{ref24} D.~S.~Petrov and G.~V.~Shlyapnikov, Phys. Rev. A {\bf 64}, 012706 (2001).



%24
\bibitem{ref22a} L.~D.~Landau and E.~M.~Lifshitz, in \textit{Quantum mechanics: Non-Relativistic Theory}, Vol. 3 (Pergamon Press, 1977) 3rd ed., Chap. 123, pp. 507--508.

%\bibitem{ref22c} L.~D.~Landau and E.~M.~Lifshitz, in \textit{Quantum mechanics: Non-Relativistic Theory}, Vol. 3 (Pergamon Press, 1977) 3rd ed., Chap. 34, pp. 114.

%25
 \bibitem{ref23} M.~Abramowitz and A.~I.~Stegun, \textit{Handbook of Mathematical Functions} (U.S. National Bureau of Standards, 1965).



%26

 \bibitem{ref25} I.~M.~Gelfand and S.~V.~Fomin, \textit{Calculus of Variations} (Dover Publications, New York, 2000).


%27
 \bibitem{ref26} W.~H.~Press, S.~A.~Teukolsky, W.~T.~Vetterling, and B.~P.~Flannery, \textit{Numerical Recipes} (Cambridge University Press, Cambridge,
1992).

%28
\bibitem{ref22h} V.~V.~Pupyshev, Phys. Atom. Nucl. {\bf 77}, 664
(2014).

%29
\bibitem{ref29} A.~N.~Kolmogorov, Ann. Math. {\bf 35}, 521 (1935).

%30
 \bibitem{ref27} E.~Haller et. al., Phys. Rev. Lett. {\bf 104}, 153203 (2010).

 \end{thebibliography}
\end{document}